\documentstyle[12pt,aaspp4]{article}

\def\etal{{et al.}}
\def\asca{{\it ASCA}}

\def\ex{{\it EXOSAT}}
\def\gi{{\it Ginga}}

\def\ros{{\it ROSAT}}
\def\pcmsq{{$\rm cm^{-2}$}}
\def\chisq{{$\chi^{2}$}}
\def\rchi{{$\chi^{2}_{\nu}$}}

\def\nh{{$N_{\rm H}$}}

\def\mdot{{$\dot{M}$}}

\begin{document}

\title{ASCA observations of Seyfert 1 galaxies: \\ I. Data Analysis, Imaging
and Timing}

\author {K. Nandra\altaffilmark{1,2}, 
I.M. George \altaffilmark{1, 3}, R.F. Mushotzky\altaffilmark{1}, 
T.J. Turner \altaffilmark{1, 3}, T. Yaqoob\altaffilmark{1, 3}}

\altaffiltext{1}{Laboratory for High Energy Astrophysics, Code 660,
	NASA/Goddard Space Flight Center,
  	Greenbelt, MD 20771}
\altaffiltext{2}{NAS/NRC Research Associate}
\altaffiltext{3}{Universities Space Research Association}

\slugcomment{To appear in {\em The Astrophysical Journal}}

\begin{abstract}

We present the first in a series of papers describing the X-ray
properties of a sample of 18 Seyfert 1 galaxies, using data obtained
by \asca. The imaging data reveal a number of serendipitous hard X-ray
sources in some source fields, but none contribute significantly to
the hard X-ray flux of the AGN. All but one of the Seyferts show
evidence for variability on timescales of minutes-hours, with the 
amplitude anti-correlated with the source luminosity, confirming
previous results. In at least 8 sources, there
is evidence that the variability amplitude below 2
keV is greater than that in the hard X-ray band, perhaps indicating
variable components other than the power-law in the soft
band. Ultra-rapid variability, implying significant power at
frequencies $> 10^{-3}$~Hz is detected in at least 5 sources, but is
difficult to detect in most cases, due to the sampling and
signal-to-noise ratio. In Mrk 766 and MCG-6-30-15 there is also
an indication that the high--frequency power--spectra are variable
in shape and/or intensity. There is similar evidence in NGC 4151, but
on longer time scales. 

\end{abstract}

\keywords{galaxies:active -- galaxies:nuclei -- X-rays: galaxies}

\section{Introduction}

\label{sec:intro}

The Ariel V and HEAO-1 surveys showed that the brightest class of
active galactic nuclei (AGN) in the hard X-ray band are nearby,
Seyfert 1 galaxies.  Whilst it is unclear whether or not such sources
are representative of the AGN phenomenon in general, the high
signal--to--noise ratio enables them to be studied them in most detail
in the X-ray band.

Long term (days to years) variability is a property of these sources
established early in the history of X-ray astronomy (e.g. Marshall,
Warwick \& Pounds 1981), with variations in amplitude of factors up to
an order of magnitude being common. Only two sources showed evidence
for more rapid variability in the HEAO-1 or {\it Einstein} data; NGC
4051 (Marshall \etal\ 1983) and NGC 6814 (Tennant \& Mushotzky
1983). In the latter case, the variability is most likely to be due to
a nearby cataclysmic variable (Madejski \etal\ 1993).
 
Long duration, uninterrupted X-ray observations of AGN were first
afforded by {\it EXOSAT}, which showed that the previous lack of
detection was due primarily to low signal-to-noise ratio and
observation duration. {\it EXOSAT}'s highly elliptical orbit
eliminated constraints due to earth occultation and SAA passage for up
to 4 days.  Examination of the light curves from short ($\sim$ 20,000
sec) observations showed that rapid variability was common, contrary
to the conclusions from low earth orbit satellites.  (e.g., Lawrence
\etal\ 1985; Pounds, Turner \& Warwick 1986).  In a study of a sample
of 48 Seyfert galaxies, at least 30\% showed clear rapid X-ray
variability (Turner 1988).

\ex\ also provided the first evenly-sampled light curves of AGN, and
therefore an opportunity to estimate the power-density spectrum (PDS)
to reasonable accuracy. The PDS can be used to identify characteristic
time scales in the variability, which might be related to physical
sizes in the source. However, early efforts showed scale-invariant
variability, with a $f^{-\alpha}$ ``red-noise'' noise spectrum,
possibly with $\alpha\sim 1$ (Lawrence \etal\ 1987; McHardy \& Czerny
1987).  Improved techniques have revealed further
information. Lawrence \& Papadakis (1993) reported on a series of
{\it EXOSAT} ``long--look'' observations and found that the power-spectrum
slopes were consistent with a mean values of $\alpha \sim 1.55\pm
0.09$.  Using different techniques, Green, McHardy \& Lehto (1993)
presented similar results for a larger sample and found $\alpha\sim
1.7\pm 0.5$. However, both sets of authors presented evidence that the
amplitude of the PDS depends on the source luminosity, confirming a
similar result based on the doubling time scale given by Barr \&
Mushotzky (1986).  Green \etal\ (1993) also reported that the
low-energy (LE) power spectra of two sources were steeper than the
medium--energy (ME) spectra.  It is still unclear whether there are
any characteristic time scales in the X-ray variability of AGN. With
the best example of a periodicity - NGC 6814 - having evaporated, the
strongest evidence is of quasi-periodic oscillations in the \ex\ power
spectra of NGC 5548 and NGC 4051 (Papadakis \& Lawrence 1993,
1995). These observations await confirmation with other
satellites. Unfortunately, all the data taken since the demise of \ex\
have been sampled unevenly.  When combined with the relatively high
Poisson noise level associated with sources of this strength, further
progress in our knowledge of the power spectra has been severely
hampered.

After introducing our sample in Section~\ref{sec:sample} and our
analysis methods in Section~\ref{sec:anal}, we discuss the \asca\
imaging data in Section~\ref{sec:im}. We demonstrate the hard X-ray
emission we describe here is consistent with an origin in a point-like
source centered on the optical galaxy and is therefore associated with
the AGN, rather than some contaminating source. The time dependence of
the X-ray emission of our Seyfert 1 galaxies is discussed in
Section~\ref{sec:lc}.
Finally, we discuss our results in Section~\ref{sec:discuss}.

\section{The Sample}

\label{sec:sample}

\begin{figure}
\plotone{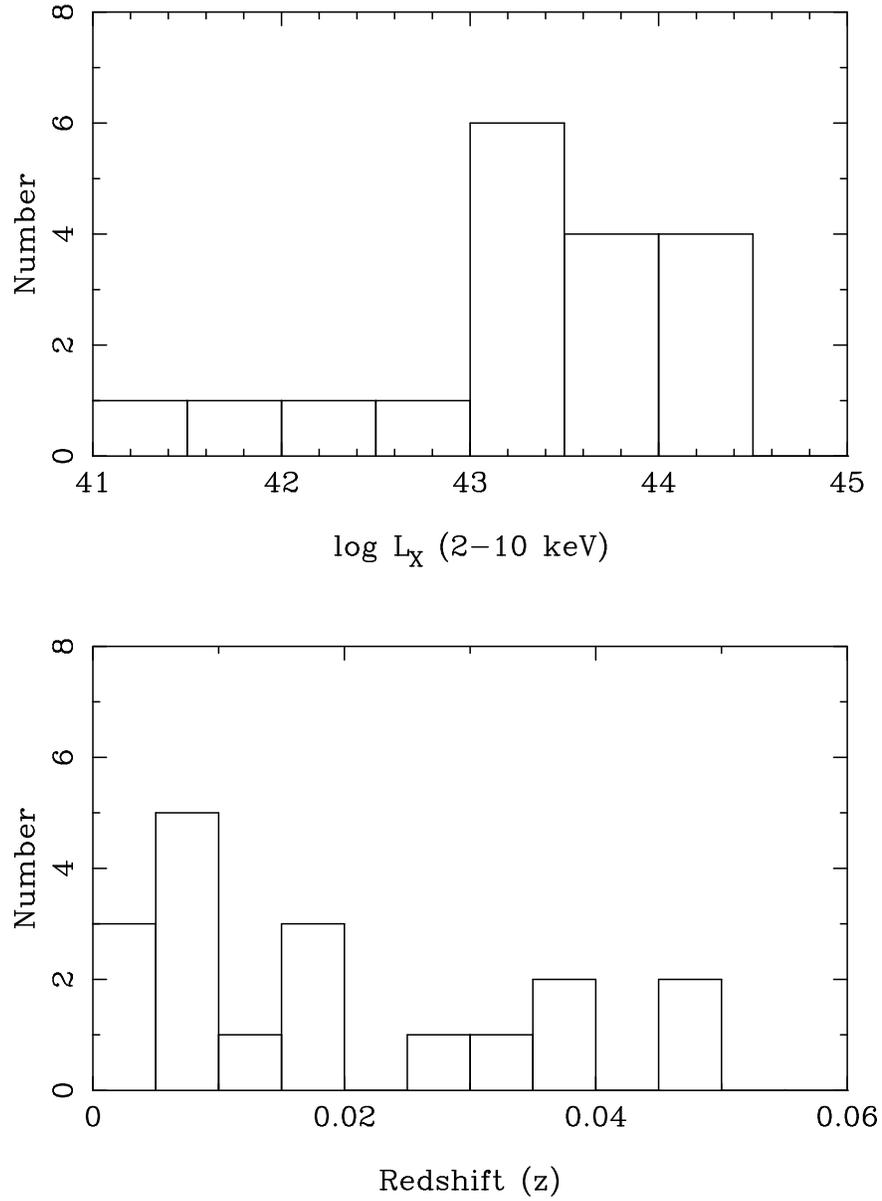}
\caption{Histograms of the X-ray luminosity (upper panel)
and redshift (lower panel) for our sample of Seyfert 1 galaxies.
The source are chosen to have redshift $z<0.05$ but
cover a wide range of $L_{\rm X}$ \label{fig:lumin}} 
\end{figure}

Here we present an analysis of \asca\ observations of Seyfert 1
galaxies performed prior to 1994 May 01.  All such observations are
now available in the public archives.  We define Seyfert 1 galaxies as
AGN at redshifts $z<0.05$, with predominantly broad lines, as
indicated by the Seyfert 1.0-1.5 classification.  We recognize that
these distinctions are somewhat arbitrary and that our sources do not
comprise a complete sample.  In the early stages of any mission, the
tendency is for the brightest X-ray sources to be observed, hence our
sample consists primarily of hard X-ray selected AGN from the HEAO-1
``Piccinotti sample'' (Piccinotti \etal\ 1982). The majority have been
studied extensively in the hard X-ray band by {\it HEAO-1} (Mushotzky
\etal\ 1984), {\it EXOSAT} (Turner \& Pounds 1989) and \gi\ (Nandra \&
Pounds 1994).  Analyses of higher-redshift and narrow-line AGN will be
presented elsewhere.  Our sample consists of 18 sources and is listed
in Table~\ref{tab:sample}. Histograms of the redshift and X-ray
luminosity distributions are shown in Fig.~\ref{fig:lumin}. The
luminosities were calculated assuming a power-law fit to the spectra,
$H_{0}=50$~km s$^{-1}$ Mpc$^{-1}$ and $q_{0}=0.5$.

As our intention is to study the mean properties of the sample we do
not include a detailed discussion of individual sources.  However, we
note that many of these observations have been published in the
literature already and the reader directed to the references noted in
Table~\ref{tab:obscat}, and those therein, for analyses of
the individual objects.

\section{Data selection and analysis}
\label{sec:anal}

Detailed information regarding \asca\ and the analysis of data
from that satellite can be found in Tanaka, Inoue \& Holt (1994), Day \etal\
1995 and references therein. Four co-aligned, grazing-incidence,
foil-mirror telescopes (Serlemitsos \etal\ 1995) are employed to
direct X-rays onto four focal-plane instruments simultaneously. There
are two CCD detectors -- the Solid-state Imaging Spectrometers (SIS; 
Gendreau 1995) and two gas--scintillation proportional--counters --
the Gas Imaging Spectrometers (GIS; Tashiro \etal\ 1995).

Our \asca\ SIS data were obtained in one of two data modes (known as
{\tt FAINT} and {\tt BRIGHT}), one of three clocking modes (1,2 or
4--CCD modes) depending on the number of CCD chips being read out and
one of three telemetry modes ({\tt LOW}, {\tt MEDIUM} and {\tt HIGH}
bit rates).  A given observation invariably contains a mixture of
these modes.  Analysis of {\tt FAINT} mode data alone allows
correction for Dark Frame Error (DFE) and ``echo'' effects
(Otani \& Dotani 1994), which
permits utilization of the maximum energy resolution of the
instruments. However, they cannot then be combined with the {\tt
BRIGHT} mode data. Thus, if signal-to-noise ratio is a more important
consideration than spectral resolution, the {\tt FAINT} data can be
converted into a format identical to the {\tt BRIGHT} mode data and
thus the two types of data can be combined.  Combining data taken in
different clocking modes can cause problems in the analysis, due to
different background and CCD zero--levels and are generally not
combined. In principle telemetry modes can be combined.

The vast majority of the GIS data described here were taken in
so-called {\tt PH} mode, which provides the maximum energy and spatial
resolution.

Our data were obtained from the US public archive (located at the
HEASARC, NASA/GSFC).  Table~\ref{tab:obscat} shows the observation
catalog.  A total of 30 observation were made of our sources, where we
define an ``observation'' as a distinct dataset within the
archive. The starting point for our analysis were the ``raw'' event
files from the database. These were created from the original
telemetry file and have been corrected to produce linearized detector
coordinates, gain corrected pulse--height values and sky
co-ordinates determined from the spacecraft attitude.  
Furthermore, the SIS {\tt FAINT} data have been
converted to both {\tt BRIGHT} mode format and an alternative form
corrected for DFE and echo effects, and compressed in PHA space,
so-called {\tt BRIGHT2} data. In cases where the majority of the data
were taken in {\tt FAINT} mode, we adopted the {\tt BRIGHT2} data.
Otherwise we combined the {\tt FAINT} mode data with the {\tt BRIGHT}
mode, to maximize the signal-to-noise ratio.  The data modes we used
are listed in column 4 of Table~\ref{tab:obscat}.  In the case of
mixed clocking modes, we chose only that mode with the highest
exposure time. This is tabulated in column 5 of
Table~\ref{tab:obscat}. The exception to this is MCG-2-58-22, where we
used the 2--CCD mode data, as even though the exposure time in 4--CCD
mode was marginally greater, 2--CCD mode is better calibrated.  We
excluded periods where telemetry saturation caused severe data
dropouts. These usually occurred with 4--CCD mode data in {\tt LOW}
bit-rate and such data were generally avoided, although in one case
saturation occurred in 1--CCD mode and another in {\tt MEDIUM}
bit-rate. Otherwise, we combined the {\tt LOW}, {\tt MEDIUM} and {\tt
HIGH} bit-rate data.

We applied data selection and cleaning algorithms using the {\tt
ascascreen} script supplied with the FTOOLS/XSELECT package
version 3.4.
Table~\ref{tab:screen} shows the selection criteria which were applied
to the datasets, unless otherwise noted in
Table~\ref{tab:obscat}. Flexible criteria (such as {\tt BR\_EARTH},
{\tt ELV} and {\tt ANG\_DIST}) were determined from an examination of
the housekeeping data.  For the SIS, the constraints on {\tt
BR\_EARTH} and {\tt ELV} were determined from an inspection of the SIS
event rates (for the on-source chip: SIS0 chip 1 or SIS1 chip 3) as a
function of these parameters. A marked increase in the event rate is
observed below the angle where significant contamination occurs.  In
cases where the the attitude ({\tt ANG\_DIST}) varied wildly (by $\gg
1$~arcmin) during the majority of the observation, that observation
was rejected.  In two observations, the attitude was relatively
stable, but the angle between the source and pointing position often
exceeded the default criterion of {\tt ANG\_DIST<0.01} used by {\tt
ascascreen} resulting in unacceptable data loss. In these cases we
allowed {\tt ANG\_DIST<0.02}.  Finally, we applied the standard
algorithm to remove ``hot'' and ``flickering'' pixels from the SIS
data.

The exposure times for the screened events files are listed in column
6 of Table~\ref{tab:obscat}.  We have excluded observations where the
exposure in either SIS was less than 10ks to ensure a similar baseline
for timing analysis and sufficient signal-to-noise ratio for spectral
analysis. Details of the the spectral analysis are presented in Nandra
\etal\ 1996b (hereafter paper II) and George \etal\ 1996 (hereafter
paper III).  This left 23 observations suitable for further analysis.

\section{Spatial analysis}
\label{sec:im}

\begin{figure}
\plotone{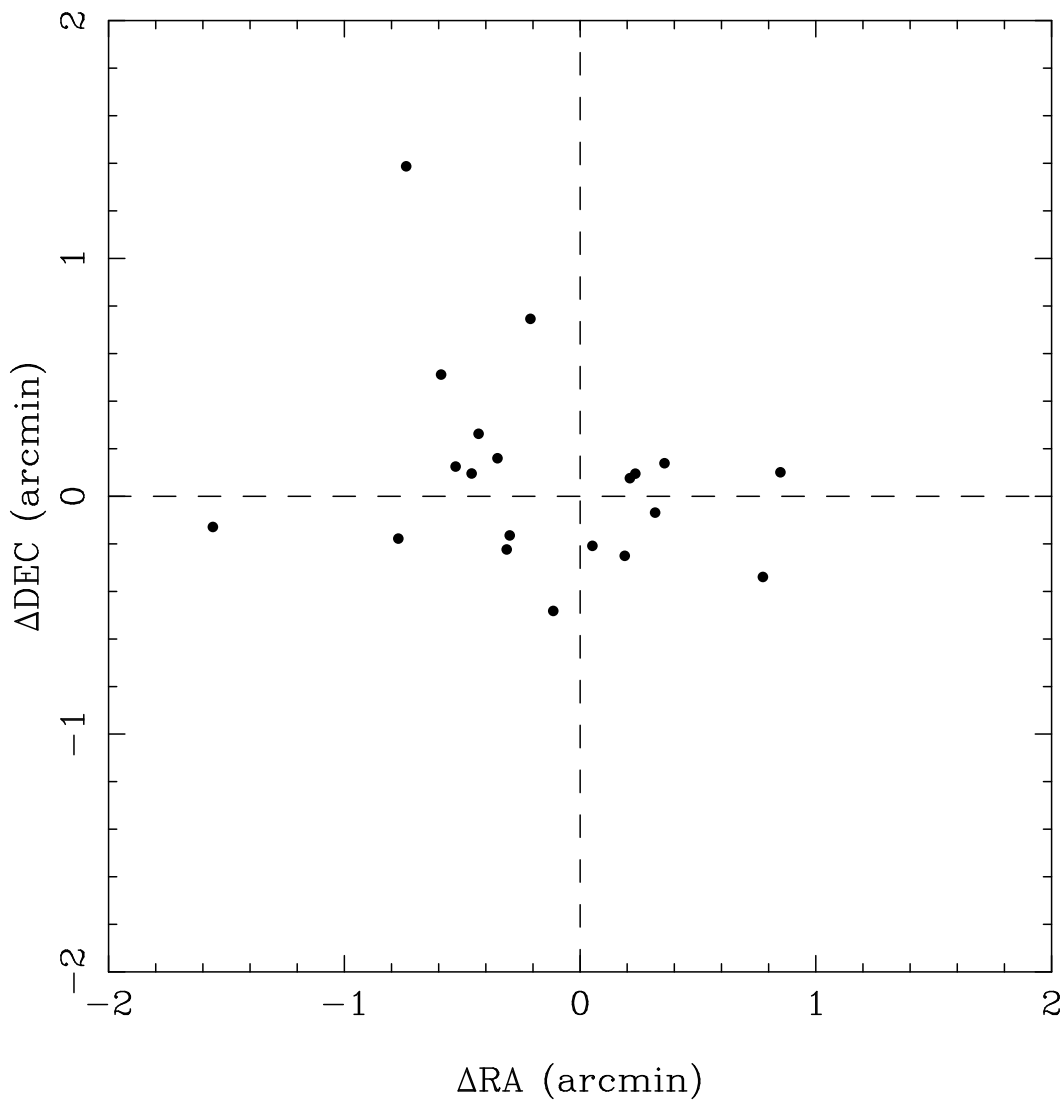}
\caption{Offset of the X-ray and optical position in Right Ascension,
versus that in Declination, both in arc minutes. 
The two positions are generally consistent within the positional
accuracy of \asca\ ($\sim 0.7$~arcmin FWHM; Gotthelf, priv. comm.).
Both NGC 4051 and Mrk 335 have
unusually large offsets, but we are confident that the AGN is the source
of the X-ray emission, as the \ros\ PSPC positions agree well with the
optical nucleus. \label{fig:offset}}
\end{figure}

After extracting the event files we accumulated and examined the
images for each instrument. The source centroids were generally within
$\sim 1$~arcmin of the optical position of the Galaxy
(Fig.~\ref{fig:offset}), which is consistent with the positional
uncertainty due to attitude reconstruction errors (Gotthelf,
priv. comm.). Mrk 335 and NGC 4051 have larger offsets ($\sim
1.5$~arcmin). Nonetheless, in both cases we are confident that the AGN
are the sources of the X-rays observed by \asca, as the \ros\ PSPC
positions, which are more accurate than those derived from the SIS,
are consistent with the optical nuclei, and show no other bright
sources within the SIS point--spread function (PSF).

We did not find any evidence for extended emission, although this is
not surprising considering the width of the PSF. Even at the distance
of the closest galaxy, the half-power diameter (HPD) corresponds to a
distance of $\sim 5$~kpc, much larger than the expected size of the
nuclear X-ray source.  Visual inspection of the SIS and GIS images
revealed six fields with evidence for sources other than the
target. These are listed in Table~\ref{tab:contam}. In all cases these
were much weaker than the Seyfert and did not cause any serious
problems in analysis. Other, even weaker sources may be present in the
images, but the presence of the central target, which is in general
very bright, makes the application of point-source searching
algorithms difficult. We have not therefore attempted any more
sophisticated spatial analysis, but naturally, we avoided the visible
contaminating sources when choosing source and background regions.

Due to small calibration differences between the SIS chips, we
restricted our analysis to the primary on-source chip 
for each SIS (SIS0 chip
1 and SIS1 chip 3).  In most cases we used a circular extraction cell for
the source region, typically $3-4$ arcmin in radius. However, in cases
where the source centroid was close to the gap between SIS chips, a
circular extraction cell resulted in an unacceptable loss of
counts. In these cases we employed rectangular extraction cells of a
similar area.  Background for the SIS was estimated using polygonal
regions at the edge of the same chip. For the GIS, the analysis was
generally simpler and in all cases we adopted a circular source region
centered on the AGN. Background counts were taken from source-free
regions.

\section{Timing analysis}
\label{sec:lc}

Initially, light curves of the source region were constructed for each
observation using the {\sc xronos} package. To increase the
signal-to-noise ratio, we combined the SIS0 and SIS1 detectors when
analyzing the SIS data; we also combined GIS2 with GIS3.  In order to
maximize the light curve data thus obtained, we initially accumulated
a light curve in 128s bins, requiring all such bins to be fully
exposed in both instruments (SIS0/SIS1 or GIS2/GIS3). This ensured
that sufficient counts were obtained in each 128s integration for
Gaussian statistics to be appropriate, even when the light curves were
split into different energy ranges (see below).  We were then able to
test for variability by means of a \chisq\ test against the hypothesis
that the flux was constant.  The reduced-\chisq\ values, \rchi, are
quoted in Table~\ref{tab:var-short} for four light curves: SIS0+SIS1
full band (0.5-10 keV; column 2), GIS2+GIS3 hard band (2-10 keV;
column 3), SIS0+SIS1 hard band (2-10 keV; column 4) and SIS0+SIS1 soft
band (0.5-2.0 keV; column 5).  We excluded NGC6814 from this analysis,
which was too weak to search for variations on these short timescales.
For the remaining sources, variations in the background have a
negligible effect on our analysis. We estimate that variability of the
SIS background would contribute less than $10^{-4}$ to the
$\sigma^{2}_{\rm RMS}$ values quoted in Table~\ref{tab:rms} and hence
a negligible effect on the \chisq\ values quoted in
Table~\ref{tab:var-short}.

For the SIS full-band light curve (column 2), 15 of the 17 objects
tested showed short-timescale variability at $>99$ per cent
confidence. The exceptions to this are Fairall-9, which showed
variability at $>95$~per cent confidence and MCG-2-58-22, which showed
no significant changes.  In almost all cases this variability is
confirmed in the GIS and in the hard and soft bands of the SIS.  We
conclude that short-timescale variability is extremely common in
Seyfert 1 galaxies, albeit at low amplitude in some cases.  

\begin{figure}
\epsscale{0.5}
\epsfysize=0.4\textwidth
\plotone{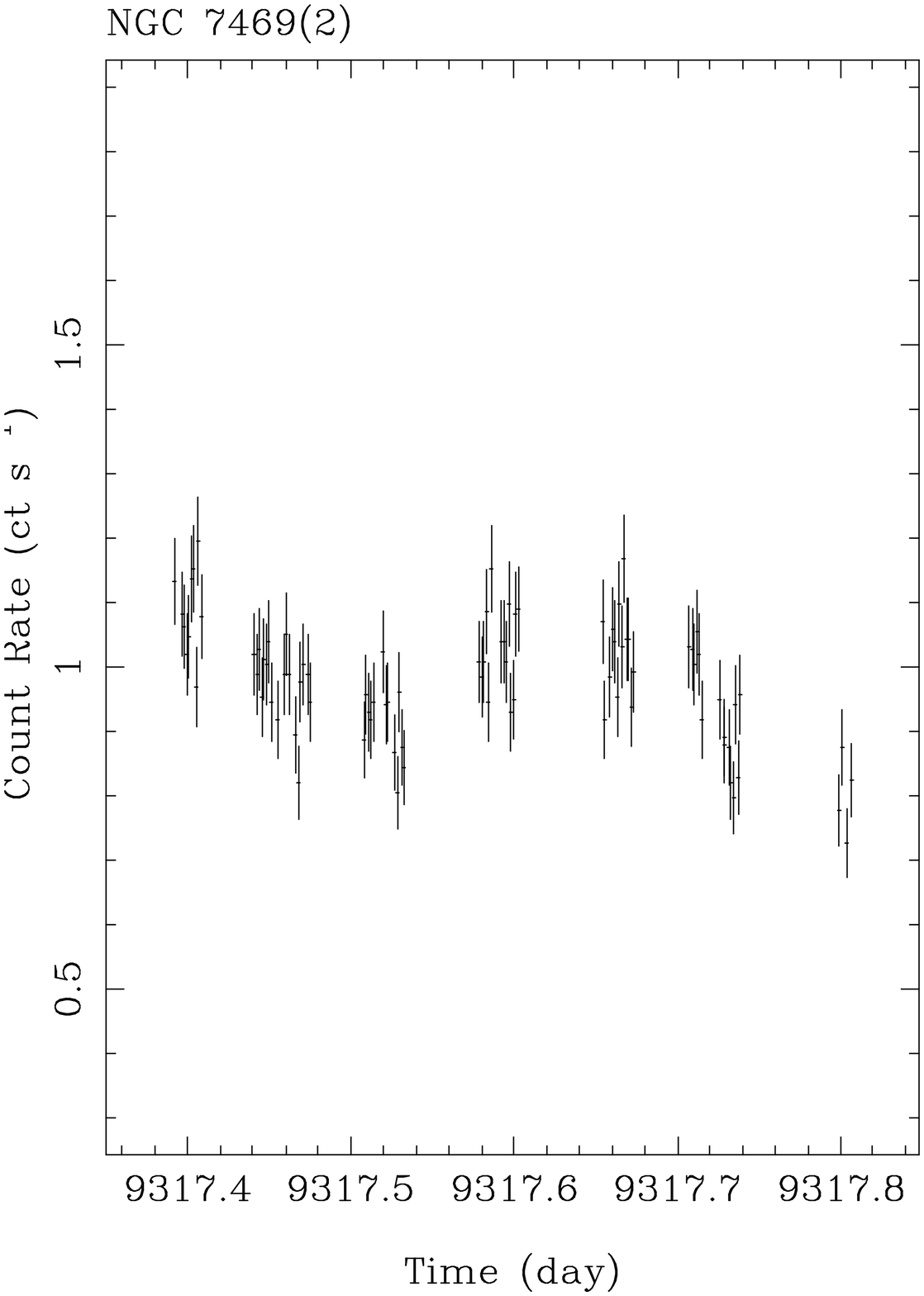}
\epsfysize=0.4\textwidth
\plotone{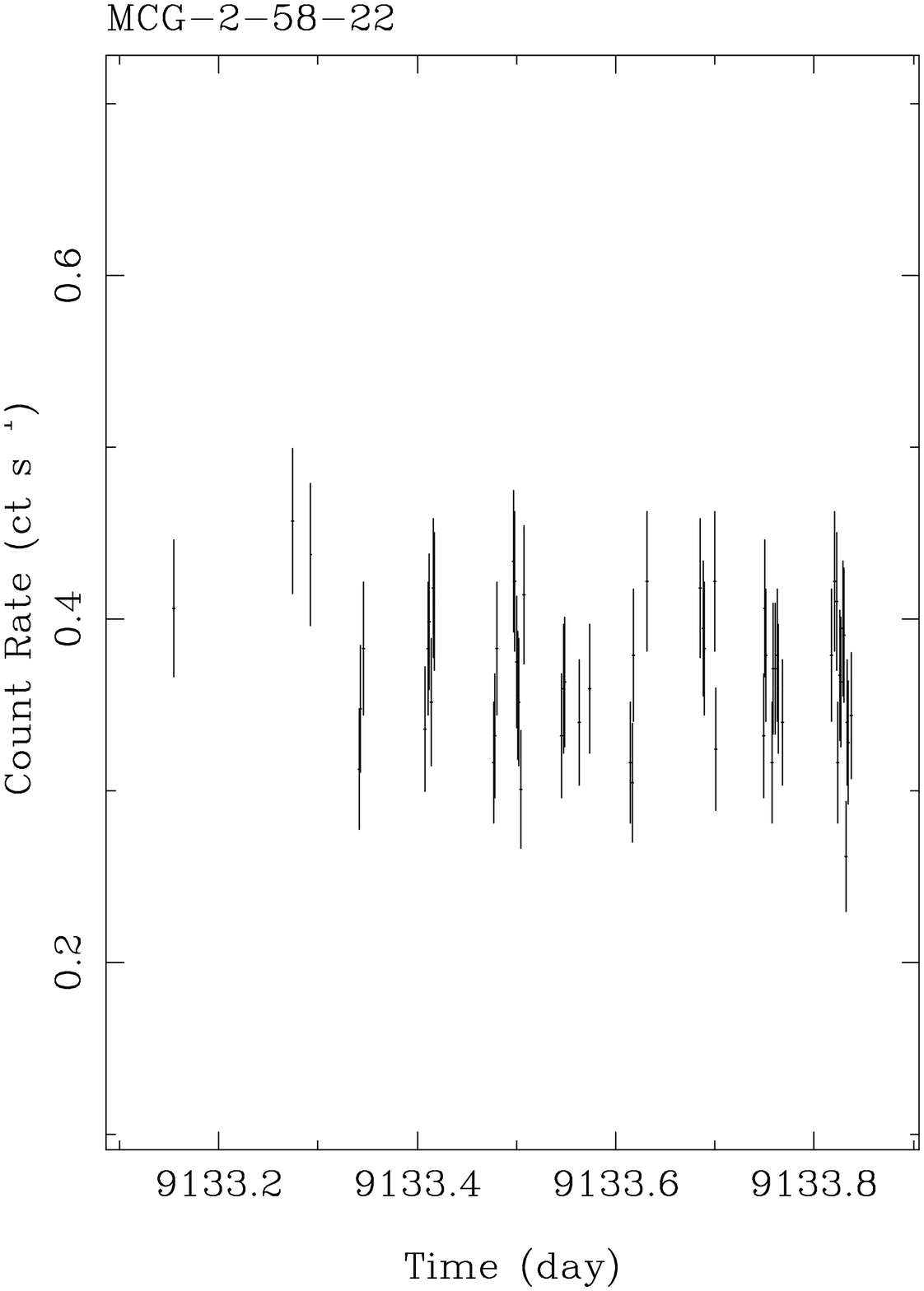}
\caption{Light curves for the sources in our sample, in 128s bins,
for the combined SIS data in the 0.4-10 keV range. The time axis
is in modified julian date, mJD-40000.
NGC 6814 has been excluded due to its low count rate.
The y-axis is scaled to the mean value, to allow comparison of
the variability amplitudes, with ymax=2*mean and ymin=mean/4,
except for NGC 4051, where the amplitude of variability
is so large that we have had to increase these limits by 50 per cent.
Most objects show variability on this timescale and also on
longer ($\sim$~hr) timescales (see Tables~\ref{tab:var-short} and
\ref{tab:var-orb}). \label{fig:lc}}  
\end{figure}

The full-band SIS light curves are shown in Fig.~\ref{fig:lc}, and
demonstrate a rich diversity of variability characteristics.  For
example, the individual sources exhibit different amplitudes of
variability. To quantify this further, we have calculated the
normalized ``excess variance'', $\sigma^{2}_{\rm RMS}$ of each light
curve.  We designate the count rates for the $N$ points in each light
curve as $X_{\rm i}$, with errors $\sigma_{\rm i}$. We further define
$\mu$ as the unweighted, arithmetic mean of the $X_{\rm i}$. Then:

$$
\sigma^{2}_{\rm RMS}=\frac{1}{N\mu^{2}}\sum_{i=1}^{N} 
[(X_{\rm i}-\mu)^2-\sigma_{\rm i}^{2}]
$$ 

The error on $\sigma^{2}_{\rm RMS}$, asymptotically for large $N$, 
is given by $s_{\rm D}/(\mu^2\sqrt{N})$ (M.G. Akritas, priv. comm) where: 

$$
s_{\rm D}^{2}=\frac{1}{N-1}\sum_{i=1}^{N} 
[(X_i-\mu)^2-\sigma_{i}^{2}]-\sigma^{2}_{\rm RMS}\mu^2
$$

i.e. the variance of the quantity $(X_i-\mu)^2-\sigma_{i}^{2}$.
These values are given in Table~\ref{tab:rms}.  As
noted by Lawrence \& Papadakis (1993) this parameter depends on
the observation length.  A more rigorous approach would
be to define the amplitude and slope of the PDS.  Unfortunately,
however, such an analysis is extremely difficult with unevenly sampled
data, as afforded by low-earth satellites such as \asca. We justify
our use of $\sigma^{2}_{\rm RMS}$ as a measure of the variability power by
noting that the $\sigma^{2}_{\rm RMS}$ is correlated with the
power-spectrum normalization in the \ex\ data (Lawrence \& Papadakis
1993) and that our observations are not radically different in
duration (Table~\ref{tab:rms}).  There are clear differences between the
sources.  Furthermore, as shown in Fig.~\ref{fig:rms}, $\sigma_{\rm
RMS}$ shows a strong anti correlation with X-ray luminosity,
confirming the previous results (see Section~\ref{sec:discuss}).  
Note that the duration of the observation, $t_{\rm D}$, is not correlated 
with luminosity.
We find $\sigma^{2}_{\rm RMS} \propto L_{\rm X}^{-0.71\pm 0.03}$, but with a
substantial scatter, particularly for the objects around 
$L_{\rm X}\sim 10^{43}$~erg s$^{-1}$. 
We note with interest, and discuss later, the fact that NGC 4151 
shows evidence for changes in $\sigma^{2}_{\rm RMS}$
at different epochs.

\begin{figure}
\epsscale{0.9}
\plotone{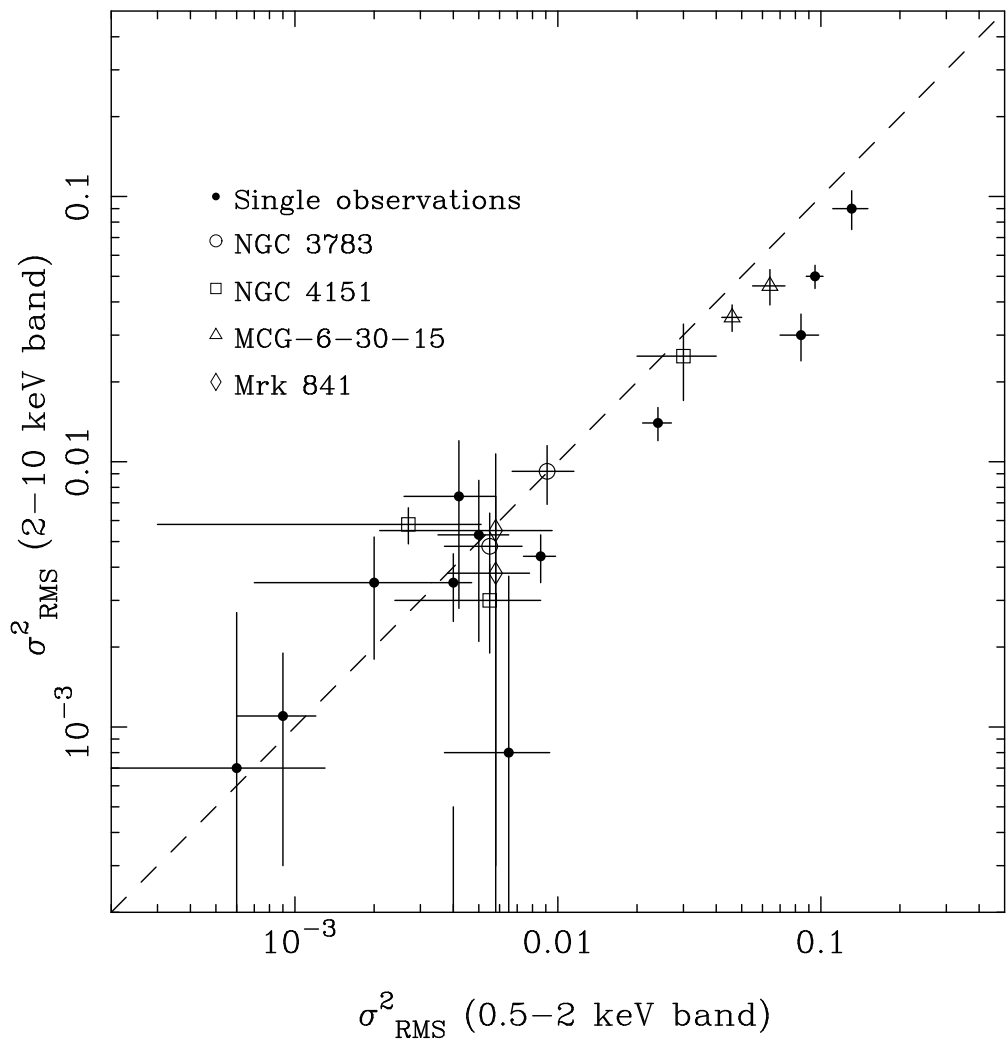}
\caption{$\sigma^{2}_{\rm RMS}$ variability amplitude
in the soft band (0.5-2 keV) versus 
that in the hard band (2-10 keV) of the SIS. The dashed
line shows a 1:1 relationship. In a number of cases,
the amplitude of variability in the soft X-ray band appears to
be greater than in the hard X-rays. This implies some spectral
variability. \label{fig:hs}}
\end{figure}

We have also employed the energy resolution of the SIS detectors to
compare $\sigma^{2}_{\rm RMS}$ for two separate energy bands. We compare
the hard band (2-10 keV) and soft band (0.5-2.0 keV) variability in
Fig.~\ref{fig:hs}. There is clearly a strong correlation. However, we
find that the amplitude of variability in the soft band is often
greater than in the hard band. This implies spectral variability for
these sources.

In Table~\ref{tab:var-orb} we show tests against the constant
hypothesis for light curves in 5760s bins (approx 1 orbit) for all our
observations.  The sources show even stronger evidence for variability
on these $\sim$hr timescales, in all instruments and energy ranges. We
were also able to test for variability in NGC 6814 on this timescale,
but no evidence for significant flux changes was found. The 
$\sigma^{2}_{\rm RMS}$ values are not quoted for these light curves as there
are generally insufficient points to make the error bars meaningful. 

\subsection{High-frequency variability}

\label{sec:var-rapid}

The high signal-to-noise ratio of our \asca\ data allows us, in
principle, to explore the PDS in a regime not accessible to \ex. Above
a frequency of $\sim 10^{-3}$~Hz, the latter data were dominated by
Poisson noise. The low background rate of the \asca\ detectors should
allow us to go beyond this point, into the regime where we might
expect cutoffs due to the fundamental size-scale of the X-ray
source. However, the uneven sampling prevents construction of a
reliable periodogram from our data. Indeed, as noted before, even the
$\sigma^{2}_{\rm RMS}$ values quoted above are susceptible to scatter
due to the sampling patterns. We have tested for the presence of very
rapid variability using light curves of the source accumulated in 32s
bins, in the SIS full band.  Even for the weakest sources, each bin
contains $\sim 20$~counts, allowing Gaussian statistics to be applied
(although once again, we excluded NGC 6814 from the analysis).  We
searched for contiguous segments with 5 or more bins in each 32s light
curve, and performed a \chisq\ test against a constant hypothesis for
each segment, flagging each time the probability of obtaining that
\chisq\ by chance was $<1$~per cent. The results are shown in
Table~\ref{tab:var-rapid}. We find at least one ``variable'' segment
(at $>99$~per cent confidence) in 10/23 observations and 9/17 of the
sources. However, given that there are typically several tens of
segments tested for each observation, the chance of spurious
detections is fairly high. We consider there to be firm evidence for
ultra-rapid variability if two or more segments show an unacceptable
\chisq, which is true in 5/17 sources. Unsurprisingly, these tend to
be the sources with the largest overall variability amplitudes.

For those sources with significant variability on these short time
scales, we also tested the hypothesis that the high-frequency PDS was
constant in shape and intensity.  We achieved this by calculating the
excess variance for contiguous segments constructed to have the same
duration. If the process producing the variability were statistically
stationary, we would then expect these $\sigma^{2}_{\rm RMS}$ values
to be constant. Given that our prescription for the error bars on the
excess variance is only valid in the limit of large $N$, we used only
segments with $>30$ points. This requirement meant that we could not
test the constancy hypothesis in NGC 4051 and NGC 3227, since there were no
such segments in the SIS light curve. The \chisq\ values and the
probabilities of obtaining them are given in columns 5 and 6 of
Table~\ref{tab:var-rapid}.  In two cases, Mrk 766 and MCG-6-30-15(2),
we find an unacceptable \chisq\ value, indicating that the high
frequency PDS in these sources is variable. Although intriguing,
we consider this result to be somewhat tentative, particularly in the
case of Mrk 766; some of the excess \chisq in that source comes from
$\sigma^{2}_{\rm RMS}$ values which are less than zero. Whilst these
are expected by chance, they may be exaggerating the significance in
this case.

\section{Discussion}
\label{sec:discuss}

We have presented \asca\ imaging and timing data for a sample of
Seyfert 1 galaxies, most of which were originally detected by
large-beam X-ray instrumentation. The \asca\ images demonstrate that,
aside from the now infamous case of NGC 6814, and despite the lack of
spatial resolution of those instruments, the Seyfert galaxies
identified in the Ariel-V and {\it HEAO-1} surveys are indeed the
sources of hard X-ray emission. There are weak contaminating sources
in some cases, but none which have a measurable impact on our current
analysis, or seem likely to have contaminated previous observations
significantly.

Our sources show variability to be very common, albeit at very low
amplitudes for some objects. This emphasizes the need for high
signal-to-noise ratio observations; this low-level variability was
missed with most previous instrumentation. The RMS variability
amplitude is strongly anti-correlated with the X-ray luminosity of the
source, but there is a substantial scatter.  Some of this could be due
to the fact that the observation durations differ, although we note
that the \ex\ data also showed a significant spread of power-spectrum
normalizations about the correlation. Were the sampling and power
spectra identical for all sources then $\sigma^{2}_{\rm RMS}$ should
be proportional to the amplitude of the PDS at a given frequency.
Therefore we compare our correlation index of $0.71$ with that given
by Green \etal\ of 0.68 and Lawrence \& Papadakis of 0.55. The fact
that these agree well, despite the expected deviations due to the
sampling patterns, supports the result that the shapes of the power
spectra are at least very similar from source to source, and that the
effects of the observation duration are not severe.

\begin{figure}
\epsscale{0.9}
\plotone{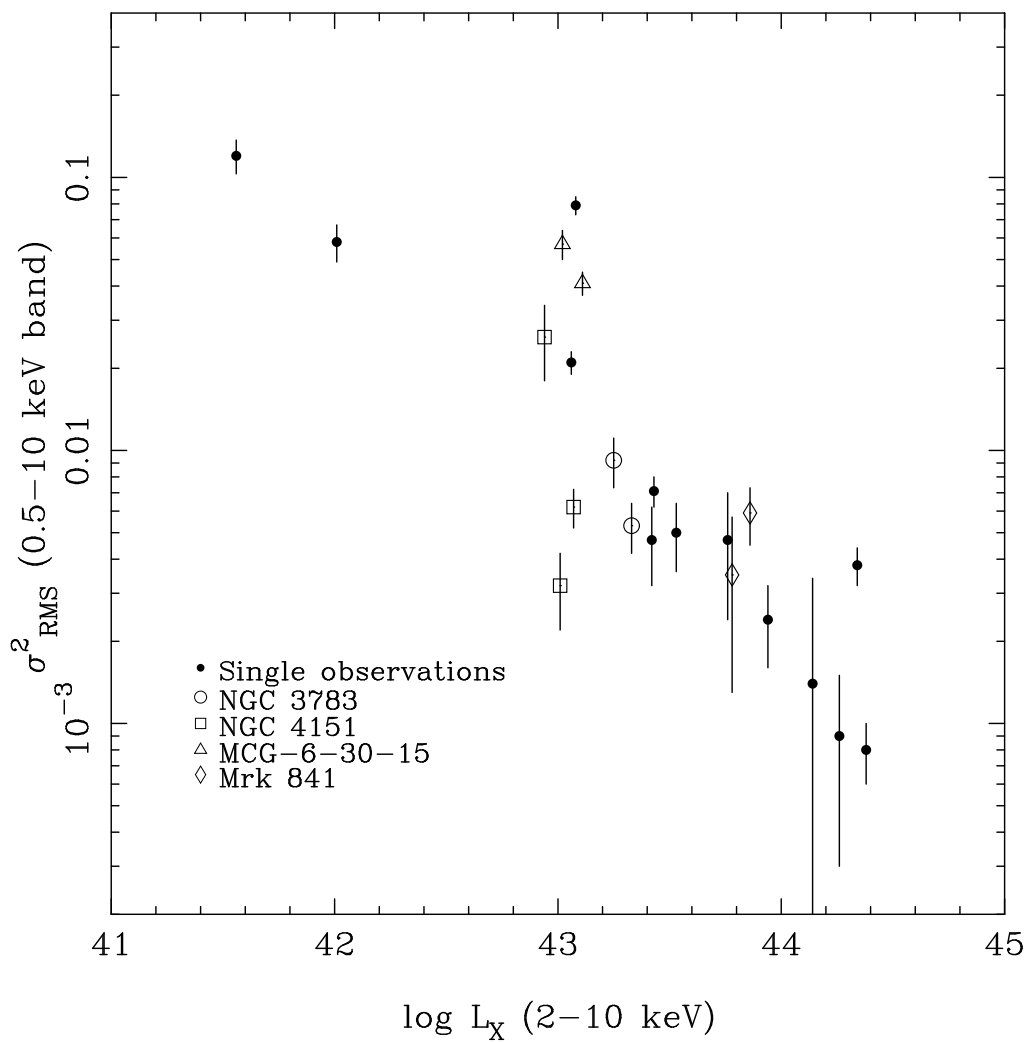}
\caption{Normalized, excess variance ($\sigma^{2}_{\rm RMS}$ 
versus luminosity. 
Where there are multiple observations
of a source, they have been plotted separately.
A highly significant 
trend of decreasing amplitude is observed with luminosity. Spearman rank
and Pearson linear correlations give significances of $>99.9$~per
cent confidence. \label{fig:rms}}
\end{figure}

It is currently unclear whether the X-ray source in Seyfert galaxies
consists of a single, coherent region or of multiple ``hot spots'',
although the lack of any obvious characteristic time scale would tend
to favor the latter (e.g., McHardy \& Czerny 1987).  In such a
scenario, one could explain an anti-correlation of $\sigma^{2}_{\rm
RMS}$ with $L_{\rm X}$ by hypothesizing that the higher luminosity
sources simply contained more hot spots, such that the large amplitude
variability was ``washed out''. We would then expect $\sigma^{2}_{\rm
RMS} \propto L_{\rm X}^{-1}$, which is somewhat steeper than the
observed correlation, although we note that the scatter evident in
Fig.~\ref{fig:rms} makes precise determination of the relation
difficult. Nonetheless, our results, and those derived from the \ex\
data, would tend to argue that the individual shots carried more power
in the higher-luminosity sources, rather than simply being more
numerous.  Alternatively, with a single X-ray-producing region, one
might hypothesize that the luminosity is related to the size of that
region. For example, if the source size is a fixed number of
gravitational radii, we expect it to be proportional to the mass of
the black hole.  With a fixed accretion rate (with respect to the
Eddington limit), \mdot, we would once again expect $\sigma^{2}_{\rm
RMS} \propto L_{\rm X}^{-1}$.  Our observed correlation would suggest
that \mdot\ increases with $L$, a view supported by some recent \ros\
and \asca\ spectral data for quasars (Stewart \etal\ 1995; Nandra
\etal\ 1996a).

The soft X-ray emission in our sources generally shows a larger
amplitude of variability than the hard X-rays on the time scales
considered.  In this case the sampling is identical for the hard and
soft X-ray light curves, so that if the power spectra were identical
in the two bands, we would not expect significant differences in
$\sigma^{2}_{\rm RMS}$. To date, the power spectra in the soft X-ray
band (e.g., the \ex\ LE) are not well known.  For two sources with
well-defined LE power spectra, NGC 4051 and MCG-6-30-15 (which are
also in our sample), it appears that the power spectra are different
in the soft and hard X-ray bands (Green \etal\ 1993; Papadakis \&
Lawrence 1995). For those sources the power spectra appear to be
steeper in the soft X-ray band, although we note that in both cases
there have been claims of quasi-periodic oscillations in the power
spectra (Papadakis \& Lawrence 1993; Papadakis \& Lawrence 1995),
making the spectral slope of the PDS much more difficult to define.
It is difficult to compare our result with theirs, as our
$\sigma^{2}_{\rm RMS}$ represents an integration of the power spectrum
over some frequency range, and cannot be easily equated to the
slope. We note, however, that, even in the soft X-ray band, the energy
spectrum is probably dominated by the power law component (see paper
III). Therefore, we would anticipate the variability characteristics
in the two bands to be similar. However, additional spectral
complexity -- and spectral variability -- in the soft X-ray band could
lead to additional variance. For example, changes in the ionization
parameter or column density of a warm absorber (e.g. Halpern 1984;
Pan, Stewart \& Pounds 1990) would have most effect in the soft X-ray
band, as would any independently-variable soft excess (e.g. Turner \&
Pounds 1988; such excesses are also clearly evident in a number of
these spectra -- see Paper III).

We find evidence that NGC 4151 shows changes in
$\sigma^{2}_{\rm RMS}$ in apparent contradiction to the suggestion that
the process producing the variability is
statistically stationary (Green \etal\ 1993; Papadakis \&
Lawrence 1995). Lawrence \& Papadakis (1993) also found changes in
$\sigma^{2}_{\rm RMS}$ between observations of NGC 4151, but as they
showed, the power spectra for those observations are consistent. 
Indeed, we find here that the longest observation of NGC 4151(2) has the
highest variance, which we expect as we are sampling the longer
time scales. Interestingly, however, a comparison of observations
(4) and (5) show the latter to have a higher variance, despite a 
shorter duration. This suggests that the shape and/or normalization
of the power spectrum changes at different epochs.
The other sources with multiple observations do not show clear evidence
for variations in $\sigma^{2}_{\rm RMS}$, when the observation
durations are taken into account. 

We found evidence for variability power on frequencies higher
than $\sim 10^{-3}$~Hz for five sources. In two cases there is also
a suggestion that the variability amplitude changes during the observation. 
This indicates that the high-frequency PDS in these sources is not
constant. A variable PDS might imply changes in fundamental time scales
in the source. At the high frequency end, this might be related to the
size-scale of the X-ray producing region(s). Our observations suggest that
it might be fruitful to pursue the study of the PDS in the 
$10^{-3}-10^{-2}$~Hz range with future instrumentation.

\acknowledgements 
We are grateful to Eric Gotthelf for his help in determining the
source positions and information regarding the positional accuracy of
the \asca\ data, Lorella Angelini for many helpful discussions, the
\asca\ team for their operation of the satellite, the \asca\ GOF at
NASA/GSFC for their assistance in data analysis, and Michael Akritas
of the SCCA operated at the Department of Statistics, Penn State
University for discussions and information on statistical
matters. This research has made use of the Simbad database, operated
at CDS, Strasbourg, France; of the NASA/IPAC Extragalactic database,
which is operated by the Jet Propulsion Laboratory, Caltech, under
contract with NASA; and data obtained through the HEASARC on-line
service, provided by NASA/GSFC. We acknowledge the financial support
of the National Research Council (KN) and Universities Space Research
Association (IMG, TJT, TY).

\clearpage

\begin{deluxetable}{l l c c c c c}

\tablecaption{The \asca\ Seyfert 1 sample. \label{tab:sample}}

\tablehead{
\colhead{Name} & \colhead{Alt. Name} & \colhead{RA (J2000)} &  
\colhead{DEC (J2000)} &  \colhead{z (V93)} & \colhead{Class} & 
\colhead{\nh (Gal)$^{a}$}
}

\startdata
Mrk 335     & PG       & 00 06 19.4 &  20 12 11  & 0.025 & 1.0 (W92) & 4.0 \nl
Fairall 9   & \nodata  & 01 23 45.7 & -58 48 21  & 0.046 & 1.0 (W92) & 3.0 \nl
3C 120      & Mrk 1506 & 04 33 11.0 &  05 21 15  & 0.033 & 1.0 (G95) & 12.3\nl
NGC 3227    & \nodata  & 10 23 30.5 &  19 51 55  & 0.003 & 1.5 (O93) & 2.2 \nl
NGC 3516    & \nodata  & 11 06 47.4 &  72 34 06  & 0.009 & 1.5 (O93) & 2.9 \nl
NGC 3783    & \nodata  & 11 39 01.7 & -37 44 18  & 0.009 & 1.2 (W92) & 8.9 \nl
NGC 4051    & \nodata  & 12 03 09.5 &  44 31 52  & 0.002 & 1.0 (O95) & 1.3 \nl
NGC 4151    & \nodata  & 12 10 32.4 &  39 24 20  & 0.003 & 1.5 (O95) & 2.1 \nl
Mrk 766     & NGC 4253 & 12 18 26.6 &  29 48 46  & 0.012 & 1.5 (W92) & 1.6 \nl
NGC 4593    & Mrk 1330 & 12 39 39.3 & -05 20 39  & 0.009 & 1.0 (W92) & 2.0 \nl
MCG-6-30-15 & \nodata  & 13 35 53.3 & -34 17 48  & 0.008 & 1.0 (G95) & 4.1 \nl
IC 4329A    & \nodata  & 13 49 19.2 & -30 18 34  & 0.016 & 1.0 (W92) & 10.4\nl
NGC 5548    & Mrk 1509 & 14 17 59.5 &  25 08 12  & 0.017 & 1.5 (O95) & 1.7 \nl
Mrk 841     & PG       & 15 04 01.1 &  10 26 16  & 0.036 & 1.5 (O95) & 2.2 \nl
NGC 6814    & \nodata  & 19 42 40.5 & -10 19 24  & 0.006 & 1.2 (W92) & 9.8 \nl
Mrk 509     & \nodata  & 20 44 09.6 & -10 43 23  & 0.035 & 1.2 (W92) & 4.2 \nl
NGC 7469    & Mrk 1514 & 23 03 15.5 &  08 52 26  & 0.017 & 1.0 (O95) & 4.8 \nl
MCG-2-58-22 & Mrk 926  & 23 04 43.4 & -08 41 08  & 0.047 & 1.2 (W92) & 3.4 \nl

\tablenotetext{a}{Galactic H{\sc i} column density
from 21cm measurements in units of $10^{20}$~\pcmsq}

\tablerefs{V93: Veron-Cetty \& Veron (1993).
G95: Giuricin, Mardirossian \& Mezzetti (1995); O93: Osterbrock \& 
Martel (1993); W92: Whittle (1992). \nh\ values are
from Elvis, Lockman \& Wilkes (1989);
Stark \etal\ (1992) or the HEASARC online service.}

\enddata
\end{deluxetable}

\clearpage

\begin{deluxetable}{lllllll}

\scriptsize
\tablecaption{\asca\ observation log. \label{tab:obscat}}

\tablehead{
\colhead{Name} & \colhead{Seq.} & 
\colhead{Date} & \colhead{CCD mode} & 
\colhead{$t_{\rm exp}$} & \colhead{Count rate} & 
\colhead{Ref}
}

\startdata

Mrk 335\tablenotemark{a}        
  & 71010000 & 93.343 & BRIGHT (2) & 19.3 & $0.584\pm 0.006$ & \nl
Fairall 9      
  & 71027000 & 93.325 & FAINT (1) & 22.0 & $1.003\pm 0.007$ & \nl
3C 120\tablenotemark{a,f}         
  & 71014000 & 94.048 & FAINT (1) & 47.4 & $1.871\pm 0.006$ & \nl
NGC 3227       
  & 70013000 & 93.128 & BRIGHT (4) & 33.4 & $0.665\pm 0.005$ & P94 \nl
NGC 3516\tablenotemark{b,f}
  & 71007000 & 94.092 & FAINT (1) & 33.7 & $2.500\pm 0.010$ & \nl
NGC 3783(1)    
  & 71041000 & 93.353 & BRIGHT (1) & 17.5 & $1.067\pm 0.008$ & G95 \nl
NGC 3783(2)    
  & 71041010 & 93.357 & BRIGHT (2) & 14.9 & $1.324\pm 0.010$ & G95 \nl
NGC 4051\tablenotemark{a}      
  & 70001000 & 93.115 & BRIGHT (4) & 27.6 & $1.063\pm 0.006$ & M94 \nl
NGC 4151(1)\tablenotemark{b,e}    
  & 70000000 & 93.144 & BRIGHT (4) & \nodata & \nodata & W94 \nl
NGC 4151(2)\tablenotemark{a}    
  & 70000010 & 93.309 & FAINT  (1) & 13.4 & $2.149\pm 0.013$ & W94 \nl
NGC 4151(3)\tablenotemark{d}    
  & 71019030 & 93.338 & BRIGHT (2) & \nodata   & \nodata  & Y95 \nl
NGC 4151(4)    
  & 71019020 & 93.339 & FAINT  (2) & 10.8 & $2.190\pm 0.015$ & Y95 \nl
NGC 4151(5)    
  & 71019010 & 93.341 & BRIGHT (2) & 11.7 & $2.479\pm 0.015$ & Y95 \nl
NGC 4151(6)\tablenotemark{e}    
  & 71019000 & 93.343 & FAINT  (2) & \nodata & \nodata & Y95 \nl
Mrk 766\tablenotemark{a}        
  & 71046000 & 93.352 & FAINT  (1) & 33.7 & $0.903\pm 0.006$ & \nl
NGC 4593\tablenotemark{b,f}       
  & 71024000 & 94.009 & FAINT  (1) & 23.1 & $1.567\pm 0.009$ & \nl
MCG-6-30-15(1) 
  & 70016000 & 93.190 & BRIGHT (2) & 29.8 & $1.961\pm 0.009$ & F94 \nl
MCG-6-30-15(2)\tablenotemark{a} 
  & 70016010 & 93.212 & BRIGHT (1) & 31.4 & $1.290\pm 0.007$ & F94 \nl
IC 4329A\tablenotemark{a,b}       
  & 70005000 & 93.227 & BRIGHT (4) & 29.1 & $2.096\pm 0.009$ & M95 \nl
NGC 5548\tablenotemark{b,c}       
  & 70018000 & 93.208 & BRIGHT (4) & 20.4 & $1.933\pm 0.010$ & M95 \nl
Mrk 841(1)\tablenotemark{g}     
  & 70009000 & 93.234 & BRIGHT (2) & 30.5 & $0.477\pm 0.004$ & G96 \nl
\tablebreak
Mrk 841(2)\tablenotemark{f}     
  & 71040000 & 94.052 & BRIGHT (2) & 20.9 & $0.355\pm 0.004$ & G96 \nl
NGC 6814(1)    
  & 70012000 & 93.124 & BRIGHT (4) & 40.3 & $0.029\pm 0.001$ & \nl
NGC 6814(2)\tablenotemark{d}    
  & 70012010 & 93.304 & BRIGHT (4) & \nodata   & \nodata   & \nl
NGC 6814(3)\tablenotemark{d}    
  & 70012020 & 93.305 & BRIGHT (4) & \nodata & \nodata   & \nl
Mrk 509        
  & 71013000 & 94.119 & FAINT  (1) & 40.1 & $2.120\pm 0.008$ & \nl
NGC 7469(1)\tablenotemark{e}    
  & 71028000 & 93.328 & FAINT  (2) & \nodata   & \nodata   & G94 \nl
NGC 7469(2)\tablenotemark{g}    
  & 71028030 & 93.330 & FAINT  (2) & 14.5 & $1.175\pm 0.009$ & G94 \nl
NGC 7469(3)\tablenotemark{e}    
  & 71028010 & 93.336 & FAINT  (2) & \nodata & \nodata  & G94  \nl
MCG-2-58-22\tablenotemark{c}    
  & 70004000 & 93.145 & FAINT  (2) & 13.3 & $0.444\pm 0.006$ & W95 \nl

\tablenotetext{a}{{\tt ELV>5}}
\tablenotetext{b}{Some data excluded due to saturation}
\tablenotetext{c}{{\tt BR\_EARTH>25} (SIS0)}
\tablenotetext{d}{Observation rejected due to bad attitude}
\tablenotetext{e}{Observation rejected due to low exposure}
\tablenotetext{f}{Affected by GIS3 BITFIX problem}
\tablenotetext{g}{{\tt ANG\_DIST<0.02}}

\tablerefs{
F94: Fabian \etal\ 1994; 
G94: Guinazzi \etal\ 1994; 
G95: George, Turner \& Netzer 1995; 
M94: Mihara \etal\ 1994; 
M95: Mushotzky \etal\ 1995;
P94: Ptak \etal\ 1994; 
W94: Weaver \etal\ 1994;
W95: Weaver \etal\ 1995; 
Y95: Yaqoob \etal\ 1995
}

\tablecomments{Column 1: Name and number of observation; Column 2: \asca\
processing sequence number; Column 3: Date of observations (year.day);
Column 4: SIS data mode and clocking mode; 
Column 5: Screened exposure time in SIS0 (ks); Column 6: Count rate in 
SIS0 (ct s$^{-1}$); Column 7: reference for previously-published \asca\ data}

\enddata
\end{deluxetable}

\clearpage

\begin{deluxetable}{ll}

\tablecaption{Selection criteria \label{tab:screen}}

\tablehead{
\colhead{Criterion} & \colhead{Description}
}
\startdata
\cutinhead{All instruments}
{\tt SAA = 0}     & Satellite outside South Atlantic Anomaly \nl
{\tt ANG\_DIST < 0.01} & Angular offset from nominal pointing position 
  (\arcdeg) \nl 
{\tt RBM\_CONT < 500} & Radiation Belt Monitor \nl
{\tt COR > 6} & Cut--off rigidity (GeV/c) \nl
\cutinhead{SIS only}
{\tt ELV > 10} & Angle from earth's limb (\arcdeg) \nl
{\tt BR\_EARTH >20} & Angle from bright earth (\arcdeg) \nl
{\tt T\_DY\_NT > 50,100,200\tablenotemark{a}} & Time after day/night 
  terminator (s) \nl
{\tt Sx\_PIXLy\tablenotemark{b} < 50,75,100\tablenotemark{a}} 
  & SIS pixel threshold \nl
\cutinhead{GIS only}
{\tt ELV>5} & Angle from earth's limb (\arcdeg) \nl

\tablenotetext{a}{For 1,2,4 CCD modes respectively} 
\tablenotetext{b}{x=SIS number; y=chip number}

\enddata
\end{deluxetable}

\clearpage

\begin{deluxetable}{l c c l l l l}

\tablecaption{Serendipitous sources \label{tab:contam}}

\tablehead{
\colhead{Field} & \colhead{RA} &  \colhead{DEC} & 
\colhead{Count Rate} &  \colhead{ID} & \colhead{Class} & 
\colhead{Det}
}

\startdata
NGC 3516 & 11 02 31 &  72 46 37 & $0.021\pm 0.002$ & MS 10590+7302 
         & QSO & GIS \nl
NGC 4151 & 12 10 25 &  39 29 24 & $0.237\pm 0.005$ & 1207+39W4 
         & BL Lac & SIS \nl 
Mrk 766  & 12 18 54 &  29 57 57 & $0.008\pm 0.001$ & \nodata 
         & \nodata & GIS  \nl
NGC 4593 & 12 40 27 & -05 13 59 & $0.010\pm 0.001$ & 1WGA J1240.4-0514 
         & \nodata & GIS \nl  
NGC 5548 & 14 18 30 &  25 10 38 & $0.076\pm 0.003$ & 1E1416.2+2525 
         & Cluster & SIS \nl    
Mrk 841  & 15 03 40 &  10 16 49 & $0.015\pm 0.001$ & \nodata
         & \nodata & GIS\tablenotemark{a} \nl   
Mrk 841  & 15 04 24 &  10 29 36 & $0.007\pm 0.001$ & PKS 1502+106 
         & HPQ & GIS \nl   

\tablenotetext{a}{Variable source}

\tablecomments{Column 1: Primary target; Column 2: Right ascension of
X-ray position of serendipitous source (J2000); 
Column 3: Declination (J2000); Column 4:
Source identification; Column 5: Instrument where detected}

\enddata
\end{deluxetable}

\clearpage

\begin{deluxetable}{lllll}

\small 
\tablecaption{\rchi/d.o.f. for 128s light curves
\label{tab:var-short}}

\tablehead{
\colhead{Name} & \colhead{SIS full} & \colhead{GIS hard} &
\colhead{SIS soft} & \colhead{SIS hard} \\
 & 
\colhead{0.5-10 keV} & \colhead{2-10 keV} &
\colhead{0.5-2 keV}  & \colhead{2-10 keV}
}

\startdata
Mrk 335        & 1.67/106 & 1.25/134 & 1.45/106 & 1.21/106  \nl
Fairall 9      & 1.23/140 & 1.07/245 & 1.12/140 & 1.08/140  \nl
3C 120         & 2.52/259 & 1.25/280 & 2.14/259 & 1.42/259  \nl
NGC 3227       & 9.14/93  & 4.06/272 & 7.71/93  & 2.94/93   \nl
NGC 3516       & 5.20/183 & 2.85/267 & 4.28/183 & 1.90/183  \nl
NGC 3783(1)    & 3.50/89  & 2.02/155 & 2.16/89  & 2.36/89   \nl
NGC 3783(2)    & 2.71/83  & 1.97/121 & 1.87/83  & 1.85/83   \nl
NGC 4051       & 20.7/83  & 7.07/184 & 17.5/83  & 4.50/83   \nl
NGC 4151(2)    & 10.7/62  & \nodata\tablenotemark{a}  & 3.64/62 
  & 8.26/62  \nl
NGC 4151(4)    & 2.81/52  & 3.00/75  & 1.49/52  & 2.45/52   \nl
NGC 4151(5)    & 4.69/64  & 4.65/92  & 1.52/64  & 13.84/64  \nl
Mrk 766        & 19.6/195 & 4.37/226 & 17.2/195 & 4.31/195  \nl
NGC 4593       & 8.01/136 & 2.76/239 & 6.74/136 & 2.47/136  \nl
MCG-6-30-15(1) & 22.4/163 & 13.4/218 & 16.9/162 & 6.30/162  \nl
MCG-6-30-15(2) & 16.4/176 & 5.37/215 & 12.3/176 & 5.40/176  \nl
IC 4329A       & 2.23/52  & 1.36/238 & 1.60/52  & 1.74/52   \nl
NGC 5548       & 3.40/19  & 1.47/208 & 3.17/19  & 1.22/19   \nl
Mrk 841(1)     & 1.67/184 & 1.20/235 & 1.48/184 & 1.14/184  \nl
Mrk 841(2)     & 1.40/95  & 1.05/123 & 1.34/95  & 1.18/95   \nl
NGC 6814(1)\tablenotemark{b}    & \nodata & \nodata & \nodata & \nodata \nl
Mrk 509        & 1.40/243 & 0.98/121 & 1.34/243 & 1.17/243  \nl
NGC 7469(2)    & 2.38/86  & 1.30/136 & 1.95/86  & 1.42/86   \nl
MCG-2-58-22    & 1.19/56  & 1.05/221 & 1.25/56  & 0.90/56   \nl
\enddata

\tablenotetext{a}{GIS exposure too low for meaningful analysis}
\tablenotetext{b}{Source too weak for analysis on this timescale}

\end{deluxetable}

\clearpage

\begin{deluxetable}{llllll}

\tablecaption{Variability amplitude ($10^{-2}\sigma^{2}_{\rm RMS}$) for 128s 
light curves \label{tab:rms}}

\tablehead{
\colhead{Name} & \colhead{SIS full} & \colhead{GIS hard} &
\colhead{SIS soft} & \colhead{SIS hard} & \colhead{$t^{a}_{\rm D}$}\\
 & 
\colhead{0.5-10 keV} & \colhead{2-10 keV} &
\colhead{0.5-2 keV}  & \colhead{2-10 keV} & \colhead{(ks)}
}

\startdata
Mrk 335        & $0.47\pm 0.15$ & $0.50\pm 0.31$ & $0.42\pm 0.16$
               & $0.74\pm 0.46$ & 50.7 \nl
Fairall 9      & $0.09\pm 0.06$ & $0.11\pm 0.15$ & $0.06\pm 0.07$
               & $0.07\pm 0.20$ & 56.7 \nl 
3C 120         & $0.38\pm 0.06$ & $0.17\pm 0.08$ & $0.40\pm 0.07$
               & $0.35\pm 0.10$ & 131.0 \nl
NGC 3227       & $5.80\pm 0.90$ & $2.80\pm 0.30$ & $8.40\pm 1.40$
               & $3.00\pm 0.60$ & 85.8 \nl
NGC 3516       & $0.71\pm 0.09$ & $0.76\pm 0.08$ & $0.86\pm 0.12$
               & $0.44\pm 0.09$ & 79.5 \nl
NGC 3783(1)    & $0.93\pm 0.19$ & $0.73\pm 0.16$ & $0.91\pm 0.24$
               & $0.92\pm 0.23$ & 38.3 \nl
NGC 3783(2)    & $0.53\pm 0.11$ & $0.64\pm 0.17$ & $0.55\pm 0.18$
               & $0.48\pm 0.16$ & 37.5 \nl
NGC 4051       & $12.0\pm 1.70$ & $5.80\pm 0.74$ & $13.1\pm 2.00$
               & $9.00\pm 1.50$ & 84.2 \nl
NGC 4151(2)    & $2.60\pm 0.80$ & \nodata\tablenotemark{b}& $3.00\pm 1.00$ 
               & $2.50\pm 0.80$ & 50.9 \nl
NGC 4151(4)    & $0.32\pm 0.10$ & $0.41\pm 0.08$ & $0.55\pm 0.31$
               & $0.30\pm 0.11$ & 29.1 \nl
NGC 4151(5)    & $0.62\pm 0.10$ & $0.67\pm 0.09$ & $0.37\pm 0.24$
               & $0.58\pm 0.09$ & 25.9 \nl
Mrk 766        & $7.90\pm 0.60$ & $5.20\pm 0.60$ & $9.50\pm 0.70$
               & $5.00\pm 0.50$ & 77.8 \nl
NGC 4593       & $2.10\pm 0.20$ & $2.20\pm 0.30$ & $2.40\pm 0.30$
               & $1.40\pm 0.20$ & 96.5 \nl
MCG-6-30-15(1) & $4.10\pm 0.40$ & $4.40\pm 0.30$ & $4.60\pm 0.40$
               & $3.50\pm 0.40$ & 87.9 \nl
MCG-6-30-15(2) & $5.70\pm 0.70$ & $4.60\pm 0.60$ & $6.40\pm 0.90$
               & $4.60\pm 0.70$ & 97.2 \nl
IC 4329A       & $0.24\pm 0.08$ & $0.10\pm 0.03$ & $0.20\pm 0.13$
               & $0.35\pm 0.17$ & 85.2 \nl
NGC 5548       & $0.47\pm 0.23$ & $0.23\pm 0.07$ & $0.65\pm 0.28$
               & $0.08\pm 0.29$ & 83.3 \nl
Mrk 841(1)     & $0.59\pm 0.14$ & $0.58\pm 0.14$ & $0.58\pm 0.20$
               & $0.38\pm 0.36$ & 79.5 \nl
Mrk 841(2)     & $0.35\pm 0.22$ & $-0.11\pm 0.34$ & $0.58\pm 0.37$
               & $0.55\pm 0.52$ & 60.9 \nl
NGC 6814(1)\tablenotemark{c}    & \nodata & \nodata & \nodata 
               & \nodata & \nodata \nl
Mrk 509        & $0.08\pm 0.02$ & $-0.02\pm 0.07$ & $0.09\pm 0.03$
               & $0.11\pm 0.08$ & 108.0 \nl
NGC 7469(2)    & $0.50\pm 0.14$ & $0.35\pm 0.19$ & $0.50\pm 0.15$
               & $0.53\pm 0.32$ & 37.4 \nl
MCG-2-58-22    & $0.14\pm 0.20$ & $0.10\pm 0.22$ & $0.40\pm 0.29$
               & $-0.24\pm 0.29$ & 83.3 \nl
\tablenotetext{a}{Duration of the observation}
\tablenotetext{b}{GIS exposure too low for meaningful analysis}
\tablenotetext{c}{Source too weak for analysis on this timescale}

\enddata

\end{deluxetable}

\clearpage

\begin{deluxetable}{lllll}

\tablecaption{\rchi/d.o.f for 5760s light curves \label{tab:var-orb}}

\tablehead{
\colhead{Name} & \colhead{SIS full} & \colhead{GIS hard} &
\colhead{SIS soft} & \colhead{SIS hard} \\
 & 
\colhead{0.5-10 keV} & \colhead{2-10 keV} &
\colhead{0.5-2 keV}  & \colhead{2-10 keV}
}

\startdata
Mrk 335        & 8.39/7  & 3.69/7  & 6.16/7  & 2.81/7  \nl
Fairall 9      & 2.11/9  & 0.93/8  & 1.84/9  & 1.25/9  \nl
3C 120         & 16.0/22 & 4.17/22 & 11.8/22 & 5.36/22 \nl
NGC 3227       & 42.8/14 & 53.1/14 & 37.0/14 & 10.2/14 \nl
NGC 3516       & 63.1/12 & 38.6/12 & 50.9/12 & 15.5/12 \nl
NGC 3783(1)    & 32.3/6  & 28.5/6  & 17.0/6  & 16.3/6  \nl
NGC 3783(2)    & 30.8/5  & 22.1/5  & 16.2/5  & 14.9/5  \nl
NGC 4051       & 119.6/12  & 54.8/12 & 97.7/12 & 23.8/12  \nl
NGC 4151(2)    & 101.0/6   & \nodata\tablenotemark{a}& 27.1/6   & 74.4/6  \nl
NGC 4151(4)    & 23.1/4  & 39.0/4  & 6.06/4  & 18.9/4  \nl
NGC 4151(5)    & 57.8/4  & 83.9/4  & 11.0/4  & 46.9/4  \nl
Mrk 766        & 233.7/13  & 40.3/13 & 212.3/13 & 37.1/13  \nl
NGC 4593       & 79.8/11 & 30.5/14 & 65.9/11 & 16.0/11 \nl
MCG-6-30-15(1) & 188.7/13  & 139.8/14  & 143.2/13 & 47.6/13  \nl
MCG-6-30-15(2) & 146.8/15  & 39.2/17 & 105.3/15 & 44.5/15  \nl
IC 4329A       & 6.29/12 & 6.30/14 & 3.59/12 & 4.22/12 \nl
NGC 5548       & 4.39/11 & 5.46/13 & 4.27/11 & 1.25/11 \nl
Mrk 841(1)     & 11.8/13 & 4.16/13 & 9.55/13 & 2.99/13 \nl
Mrk 841(2)     & 5.04/10 & 1.51/10 & 4.78/10 & 1.59/10 \nl
NGC 6814(1)    & 0.97/20 & 0.92/23 & 1.39/20 & 0.63/20 \nl
Mrk 509        & 7.45/18 & 1.92/6  & 5.38/18 & 3.43/18 \nl
NGC 7469(2)    & 16.8/6  & 11.2/6  & 11.8/6  & 6.05/6  \nl
MCG-2-58-22    & 1.55/10 & 1.57/14 & 1.38/10 & 1.95/10 \nl

\tablenotetext{a}{GIS exposure too low for meaningful analysis}

\enddata
\end{deluxetable}

\clearpage

\begin{deluxetable}{lccccc}

\tablecaption{Rapid variability; 32s bins \label{tab:var-rapid}}

\tablehead{
\colhead{Name} & \colhead{No. of} & \colhead{Longest} & \colhead{Variable} &
\colhead{\rchi$^{a}$/d.o.f.} & \colhead{$prob^{b}_{var}$} \\ & 
\colhead{Segments} & \colhead{(s)} & \colhead{Segments} & 
\colhead{($\sigma^{2}_{\rm RMS}$)} & \colhead{($\sigma^{2}_{\rm RMS}$)} 
}

\startdata
Mrk 335        & 25  & 2112 & 0 & \nodata & \nodata \nl
Fairall 9      & 30  & 1984 & 0 & 0.42/4  & 0.79 \nl
3C 120         & 69  & 1632 & 0 & 1.25/15 & 0.22 \nl
NGC 3227       & 55  & 864  & 2 & \nodata & \nodata \nl
NGC 3516       & 37  & 1824 & 2 & 1.06/12 & 0.39 \nl
NGC 3783(1)    & 26  & 1248 & 1 & \nodata & \nodata \nl
NGC 3783(2)    & 18  & 1312 & 0 & \nodata & \nodata \nl
NGC 4051       & 46  & 864  & 2 & \nodata & \nodata \nl
NGC 4151(2)    & 22  & 1152 & 0 & \nodata & \nodata \nl
NGC 4151(4)    & 16  & 1600 & 0 & \nodata & \nodata \nl
NGC 4151(5)    & 19  & 1312 & 0 & \nodata & \nodata \nl
Mrk 766        & 37  & 1760 & 3 & 2.99/13 & $<10^{-3}$ \nl
NGC 4593       & 24  & 1632 & 0 & 0.88/9  & 0.54 \nl
MCG-6-30-15(1) & 45  & 1248 & 5 & 5.95/5  & $<10^{-4}$ \nl
MCG-6-30-15(2) & 49  & 1504 & 2 & 0.88/8  & 0.54 \nl
IC 4329A       & 31  & 928  & 0 & \nodata & \nodata \nl
NGC 5548       & 19  & 352  & 0 & \nodata & \nodata \nl
Mrk 841(1)     & 39  & 2112 & 1 & \nodata & \nodata \nl
Mrk 841(2)     & 33  & 1120 & 0 & \nodata & \nodata \nl
NGC 6814(1)    & \nodata & \nodata & \nodata & \nodata & \nodata \nl
Mrk 509        & 62  & 1632 & 1 & 1.05/13 & 0.40 \nl
NGC 7469(2)    & 19  & 1536 & 0 & 0.70/ 4  & 0.59 \nl
MCG-2-58-22    & 25  & 1280 & 1 & \nodata & \nodata \nl

\tablenotetext{a}{For a test of the hypothesis that $\sigma^{2}_{\rm RMS}$
was constant. We quote values only for those observations with
$>4$ segments with $>30$ points and used identical sampling 
for each segment (see text)}
\tablenotetext{b}{Probability of obtaining this \chisq\ by chance}
\enddata
\end{deluxetable}

\clearpage

\begin{figure}
\epsscale{0.5}
\epsfysize=0.4\textwidth
\plotone{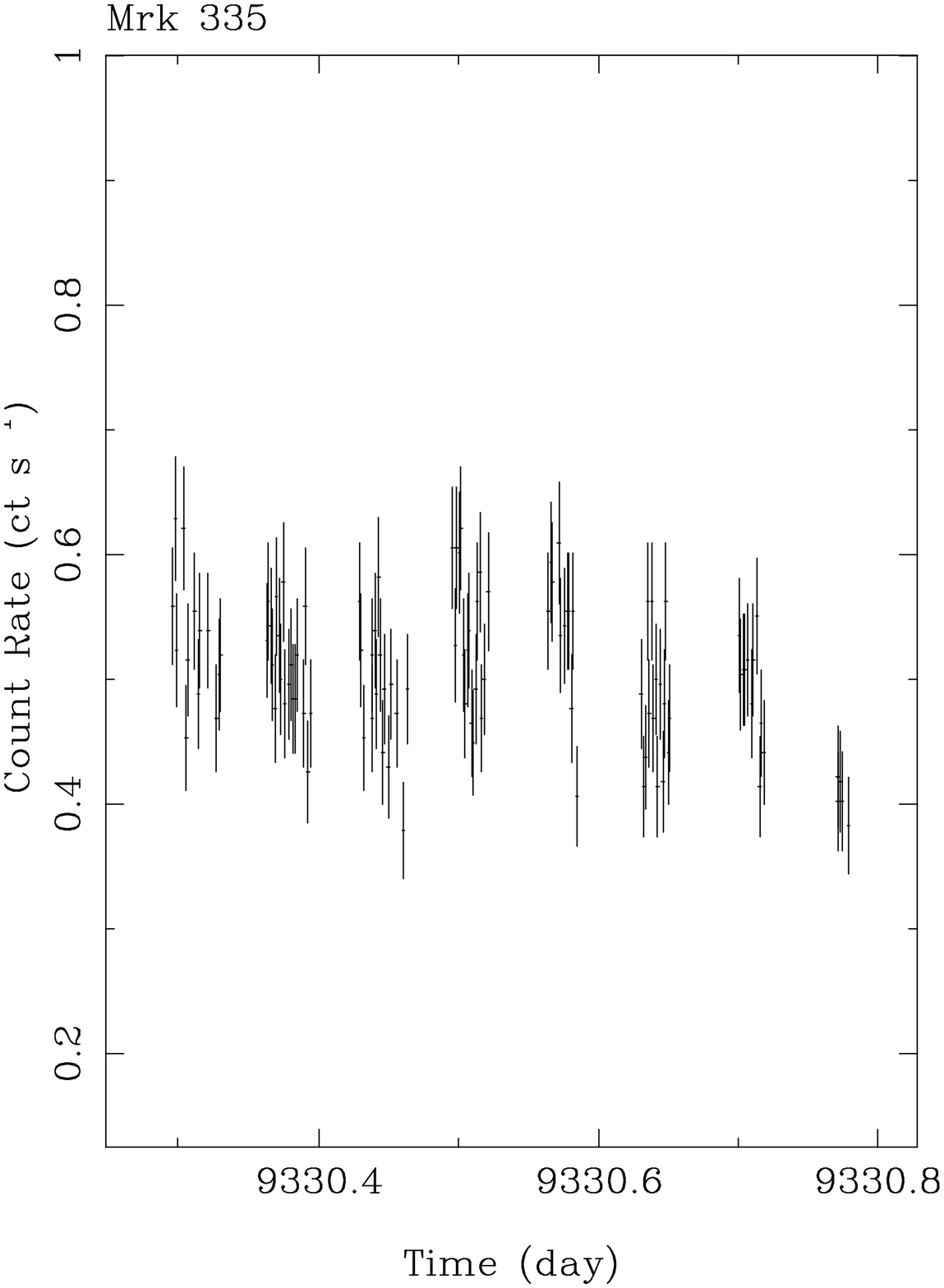}
\epsfysize=0.4\textwidth
\plotone{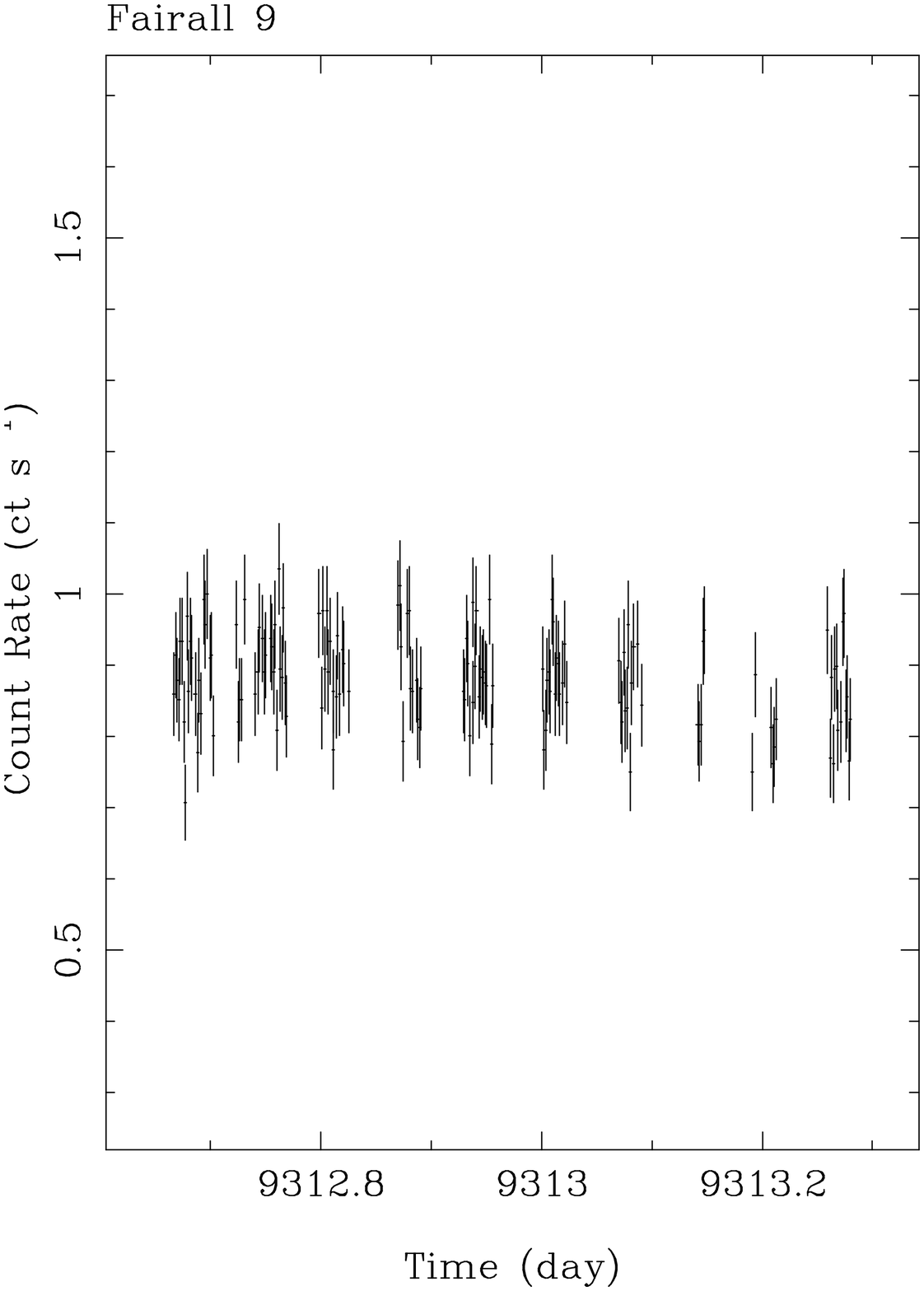}
\end{figure}

\begin{figure}
\epsscale{0.5}
\epsfysize=0.4\textwidth
\plotone{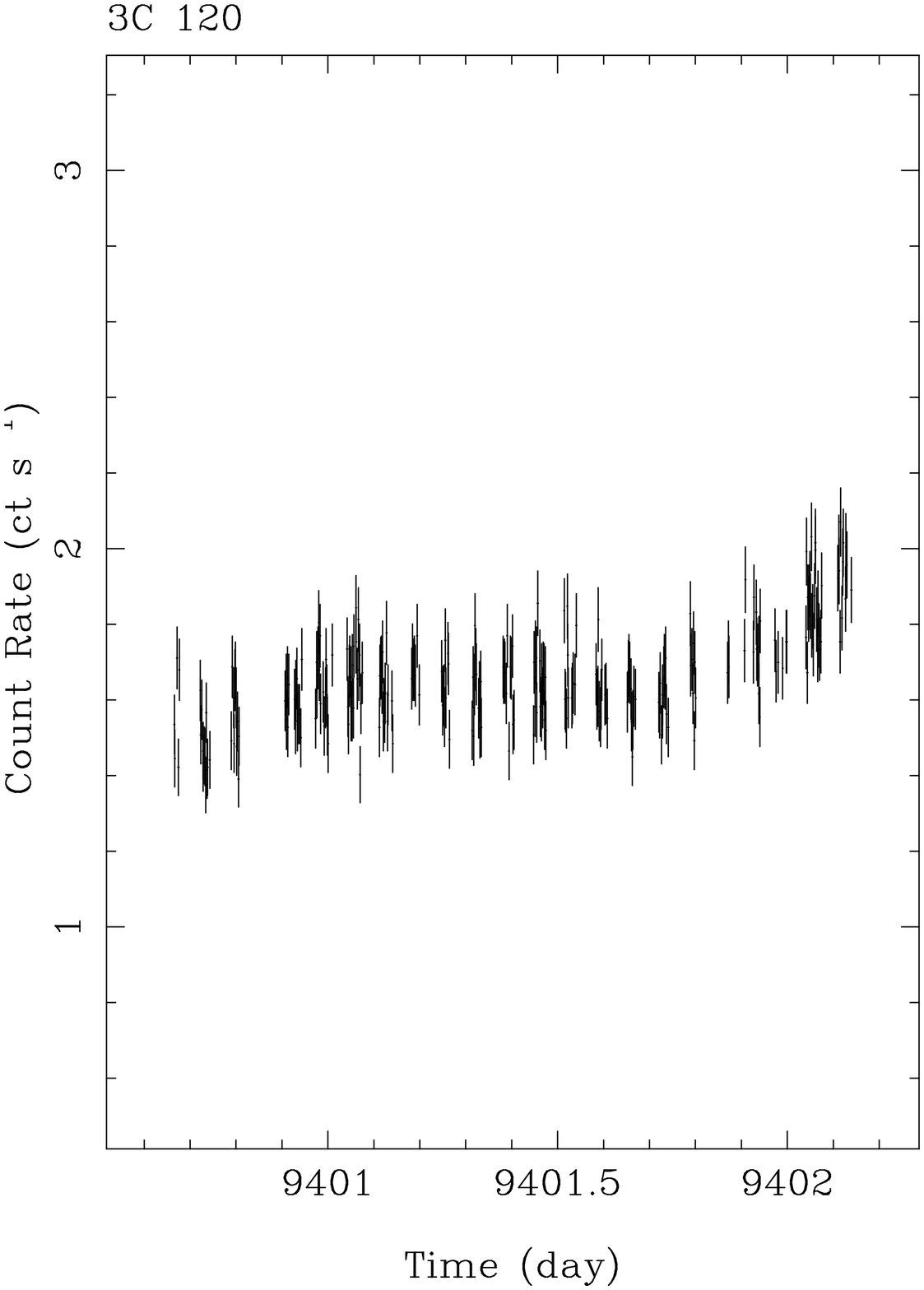}
\epsfysize=0.4\textwidth
\plotone{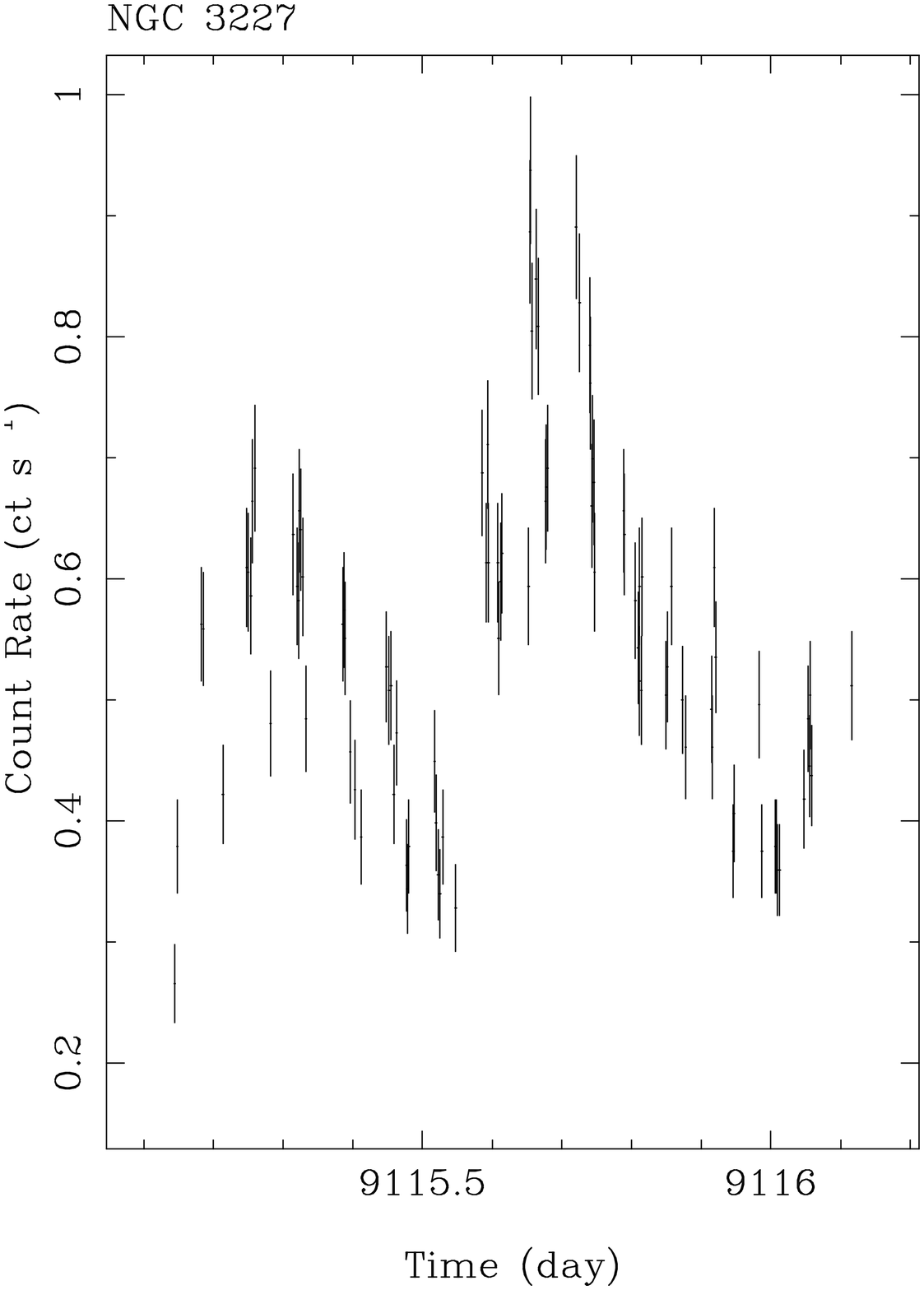}
\end{figure}

\begin{figure}
\epsscale{0.5}
\epsfysize=0.4\textwidth
\plotone{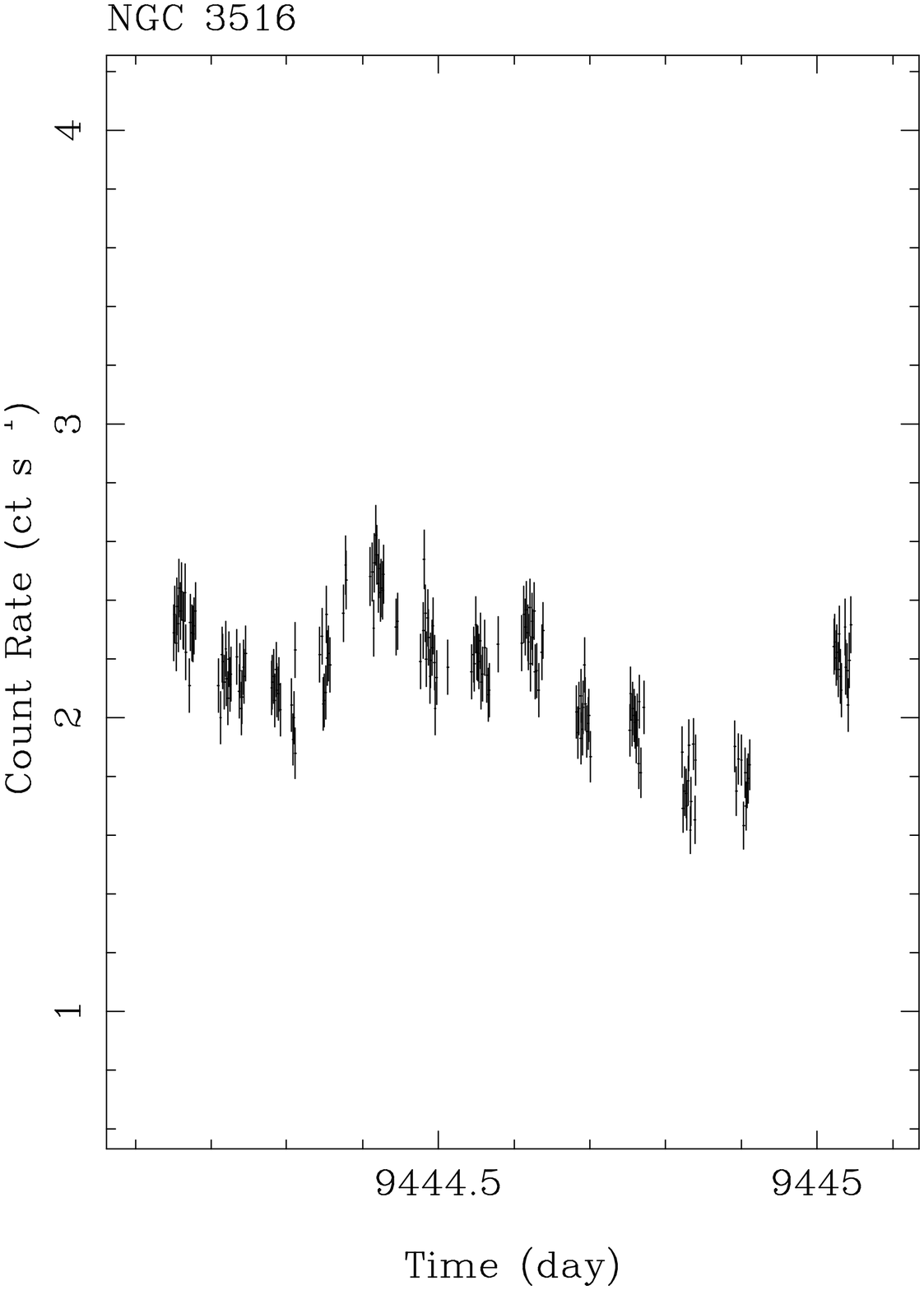}
\epsfysize=0.4\textwidth
\plotone{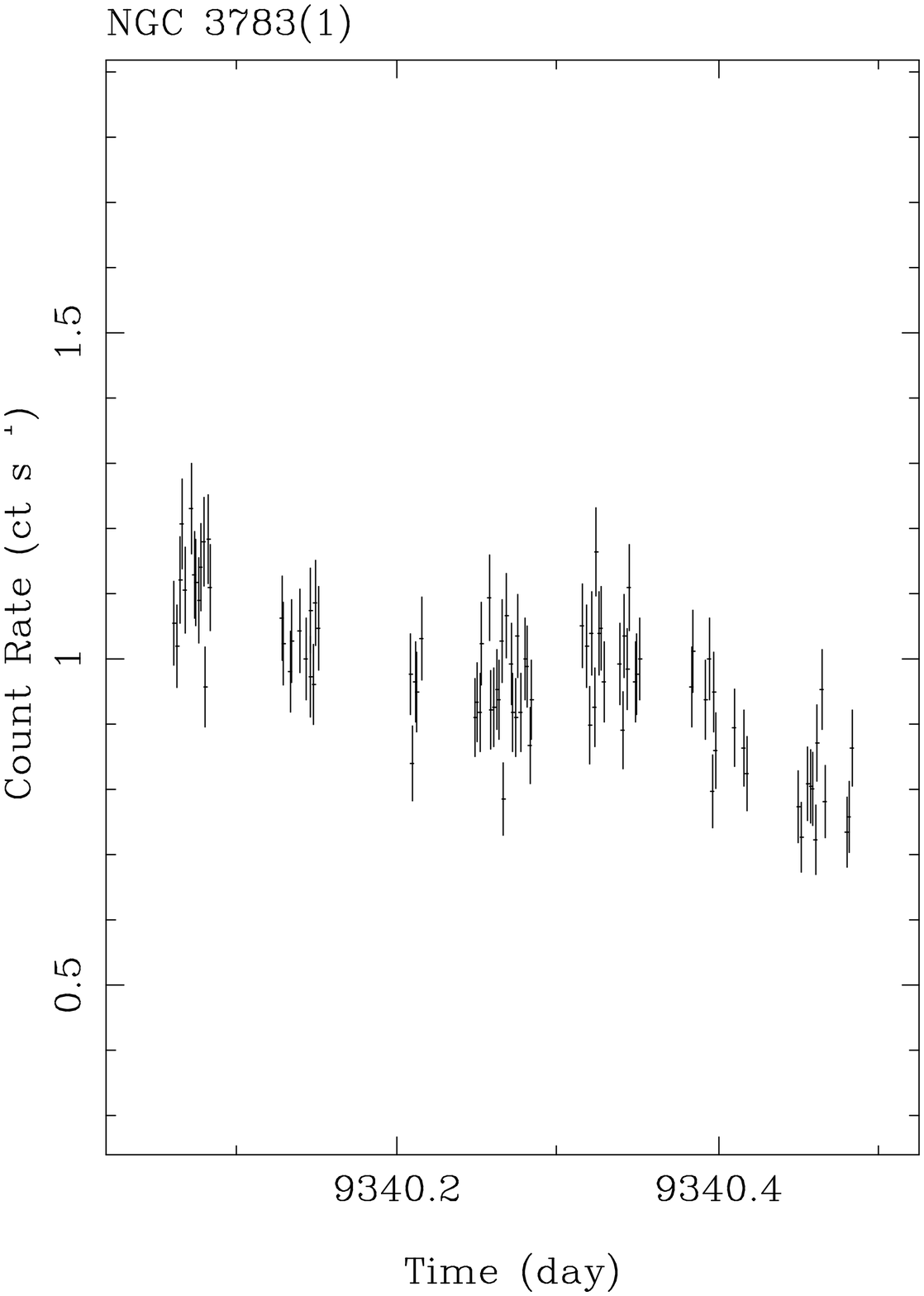}
\end{figure}

\begin{figure}
\epsscale{0.5}
\epsfysize=0.4\textwidth
\plotone{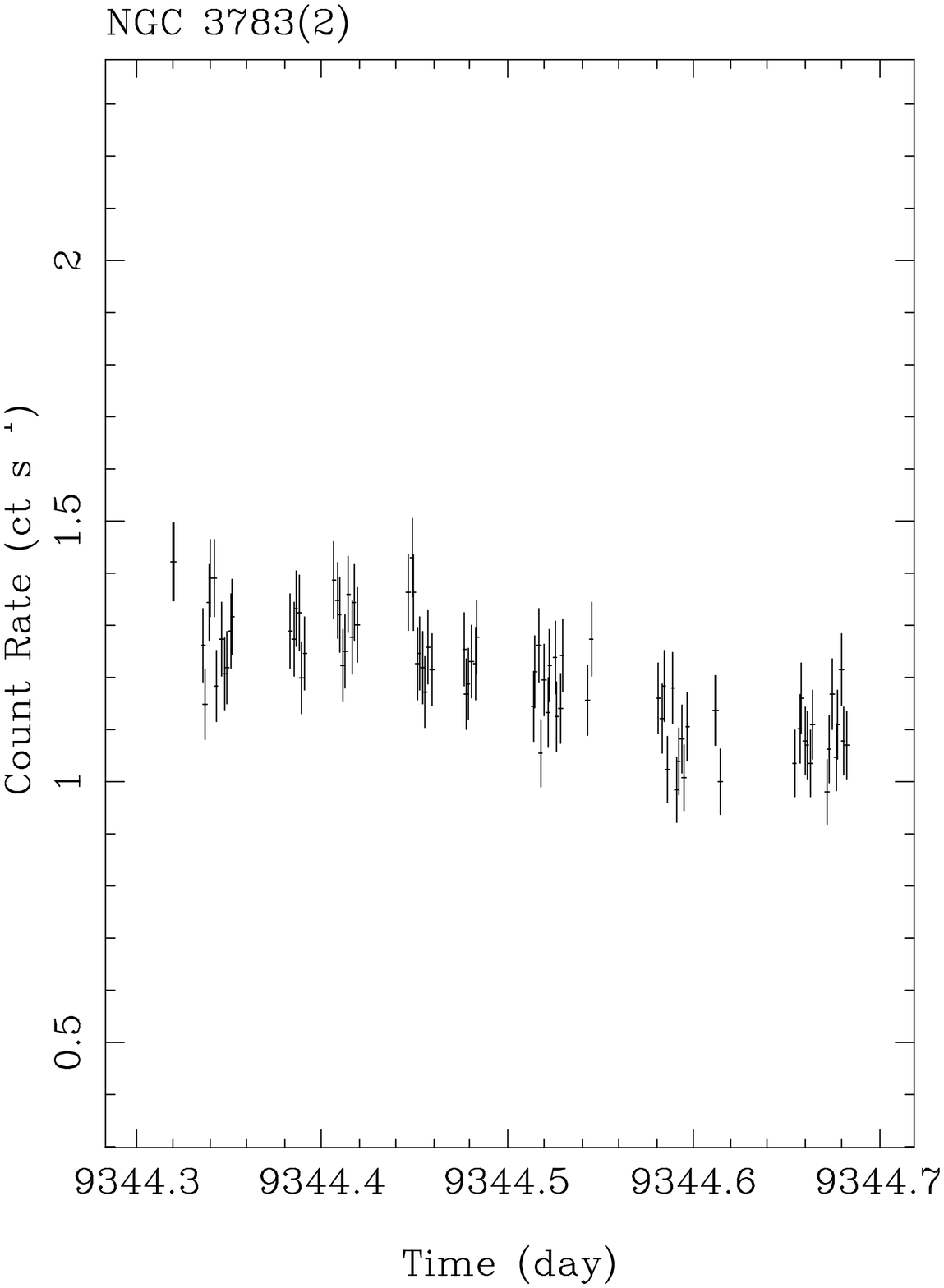}
\epsfysize=0.4\textwidth
\plotone{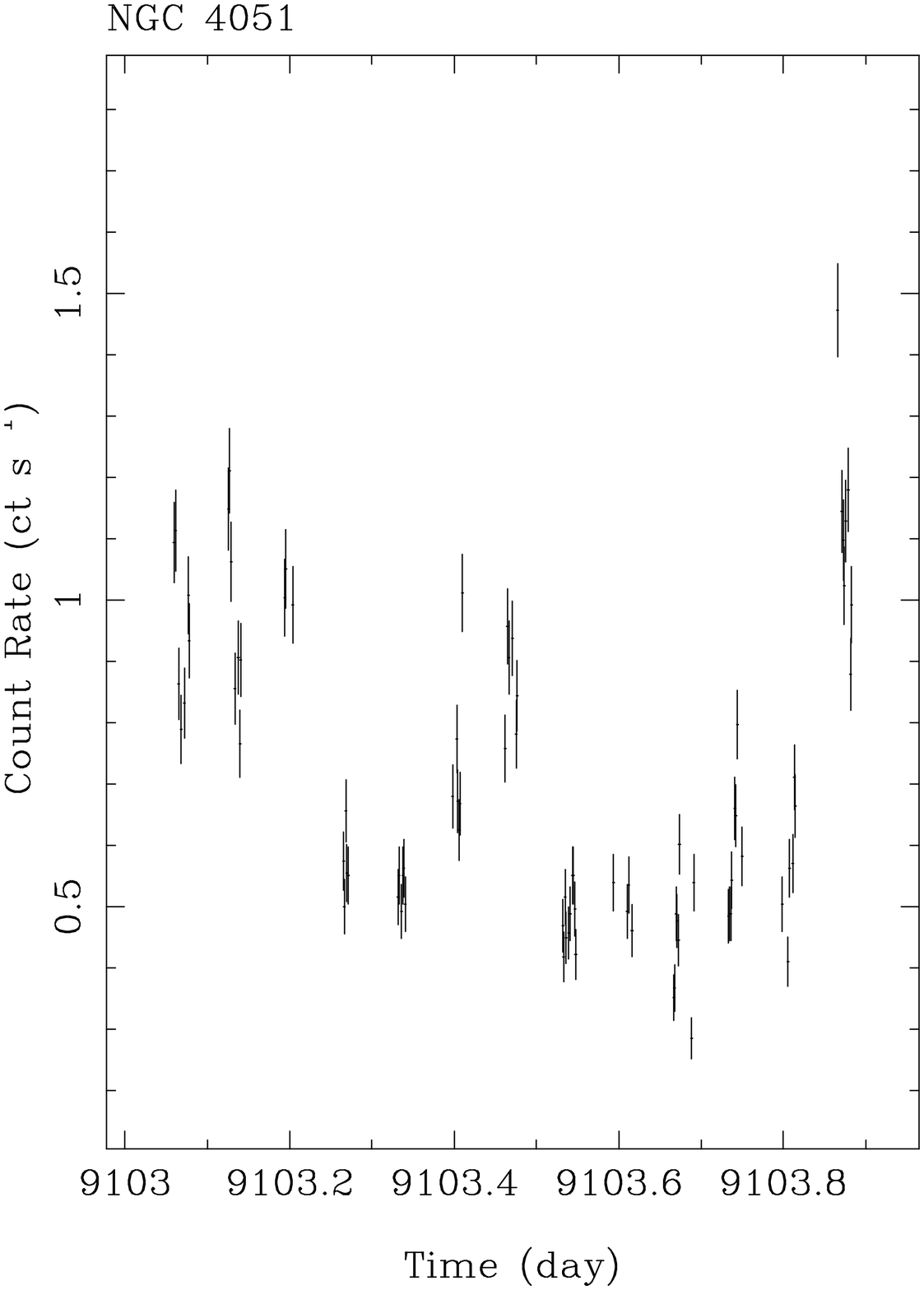}
\end{figure}

\begin{figure}
\epsscale{0.5}
\epsfysize=0.4\textwidth
\plotone{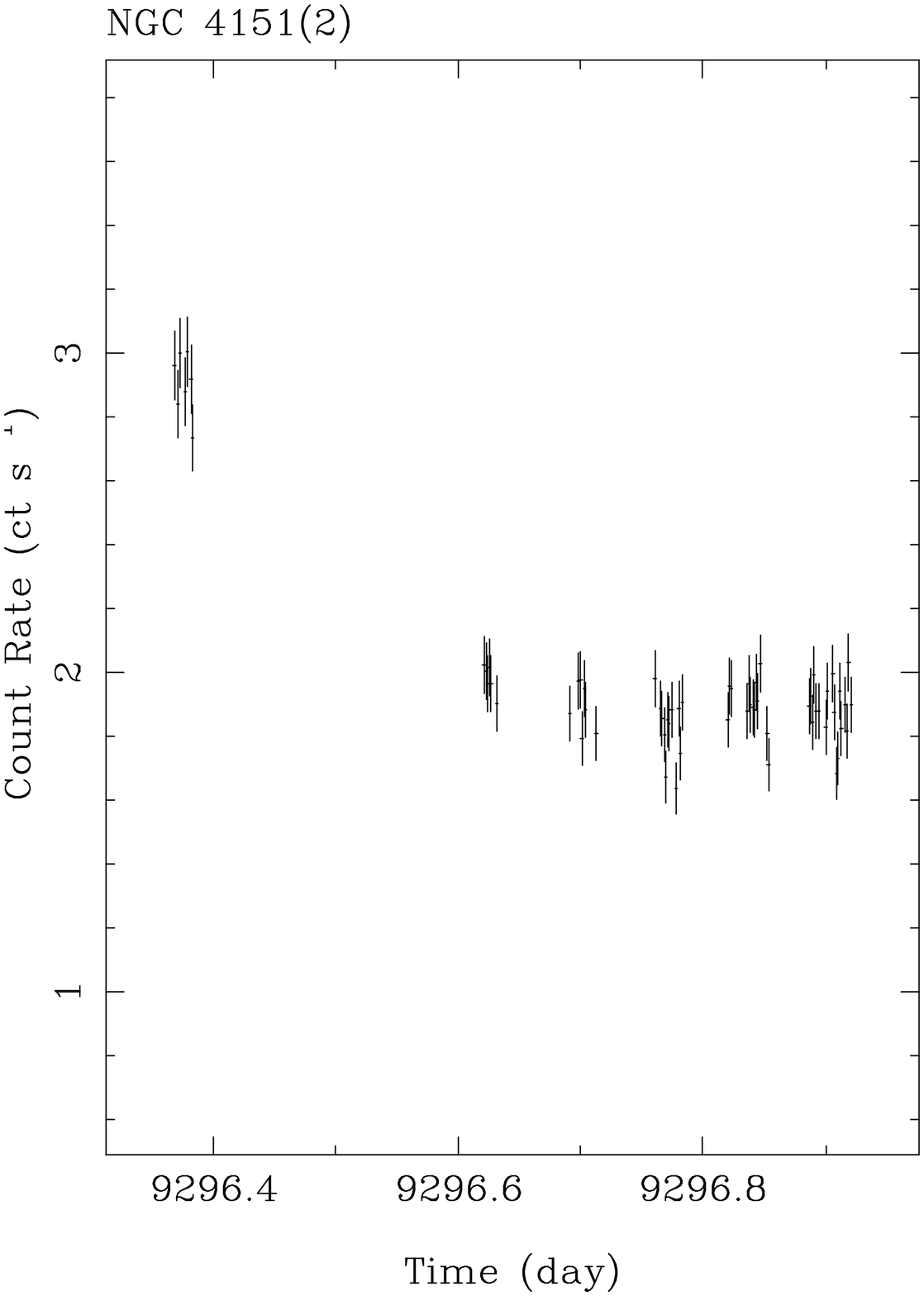}
\epsfysize=0.4\textwidth
\plotone{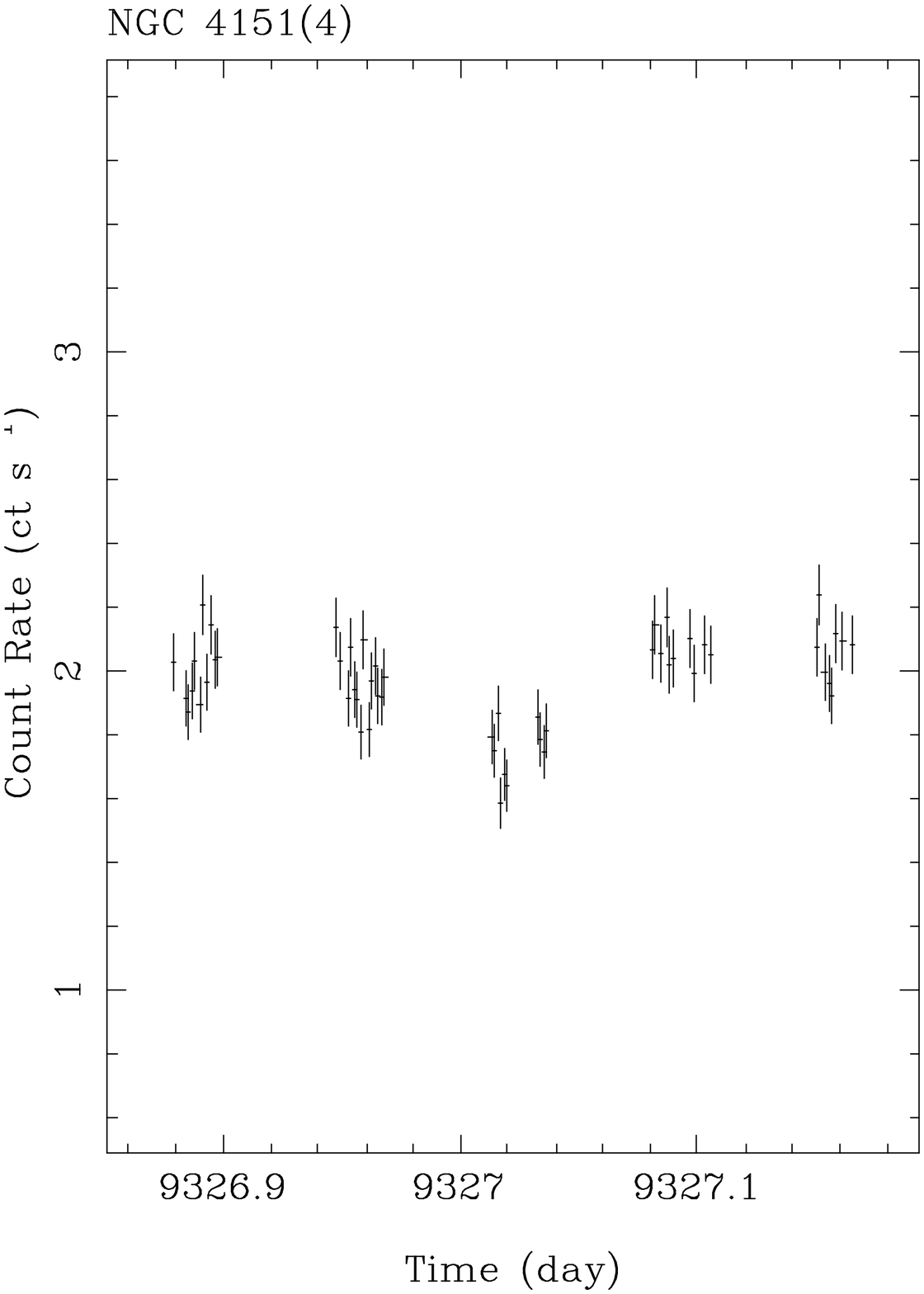}
\end{figure}

\begin{figure}
\epsscale{0.5}
\epsfysize=0.4\textwidth
\plotone{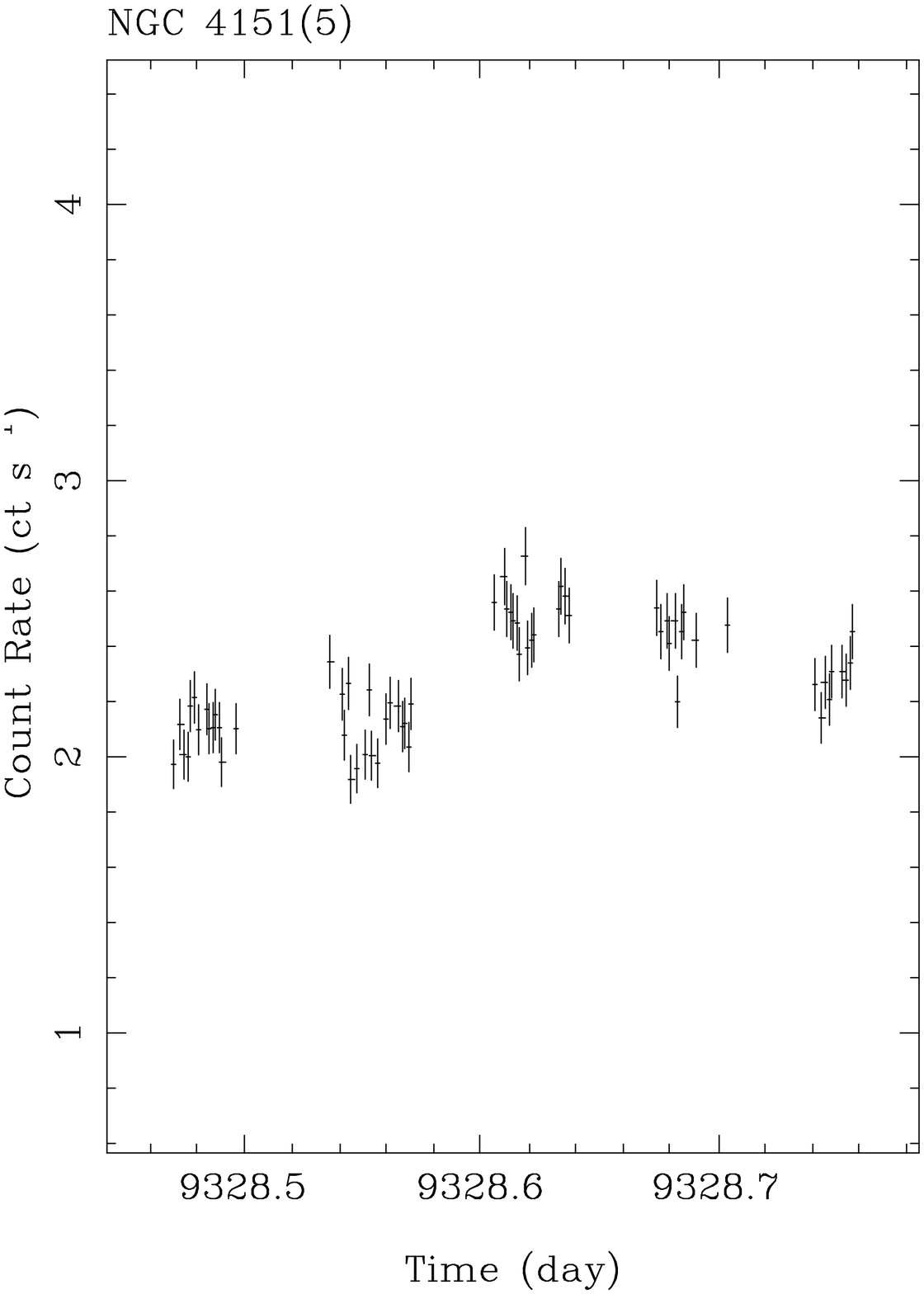}
\epsfysize=0.4\textwidth
\plotone{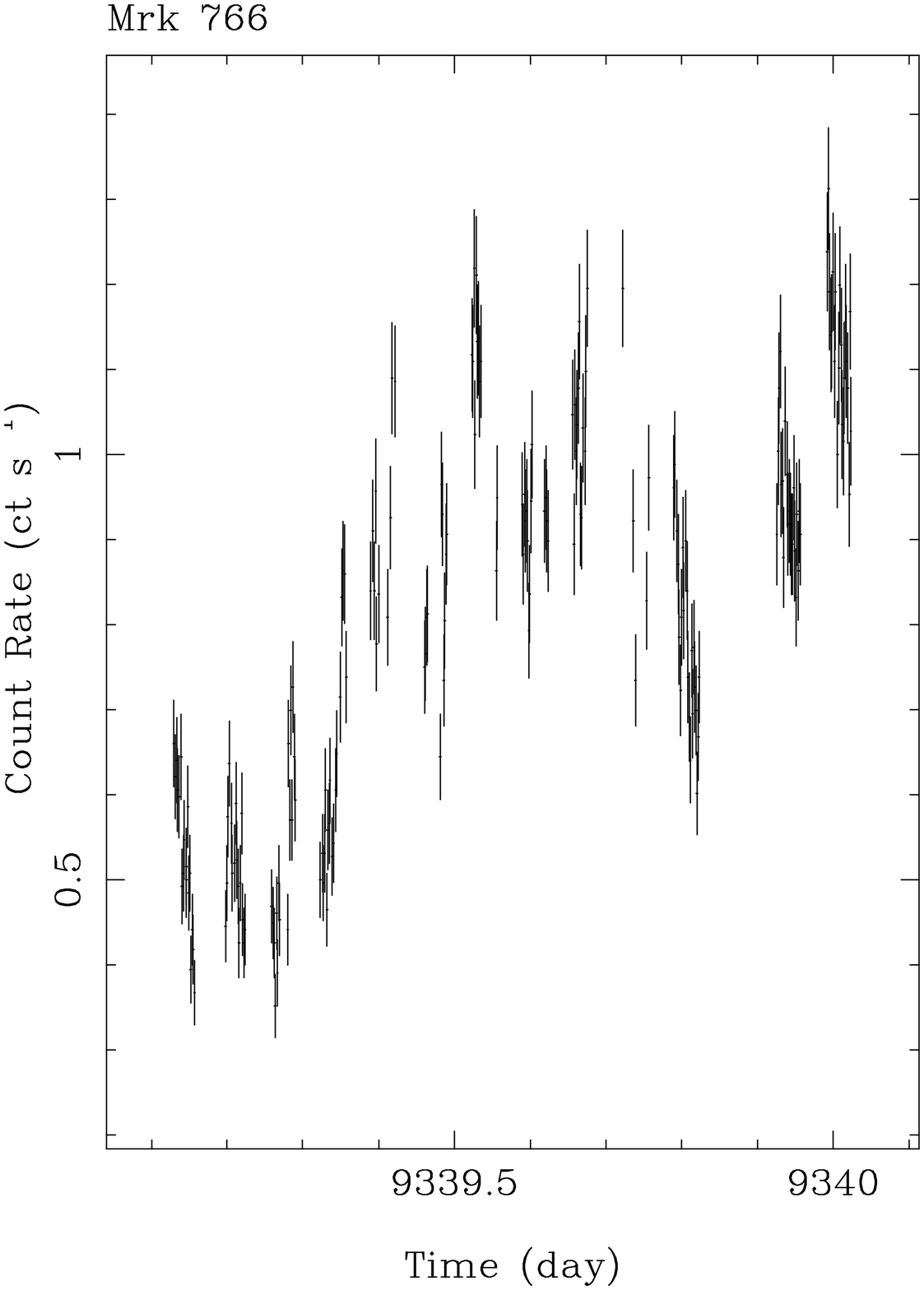}
\end{figure}

\begin{figure}
\epsscale{0.5}
\epsfysize=0.4\textwidth
\plotone{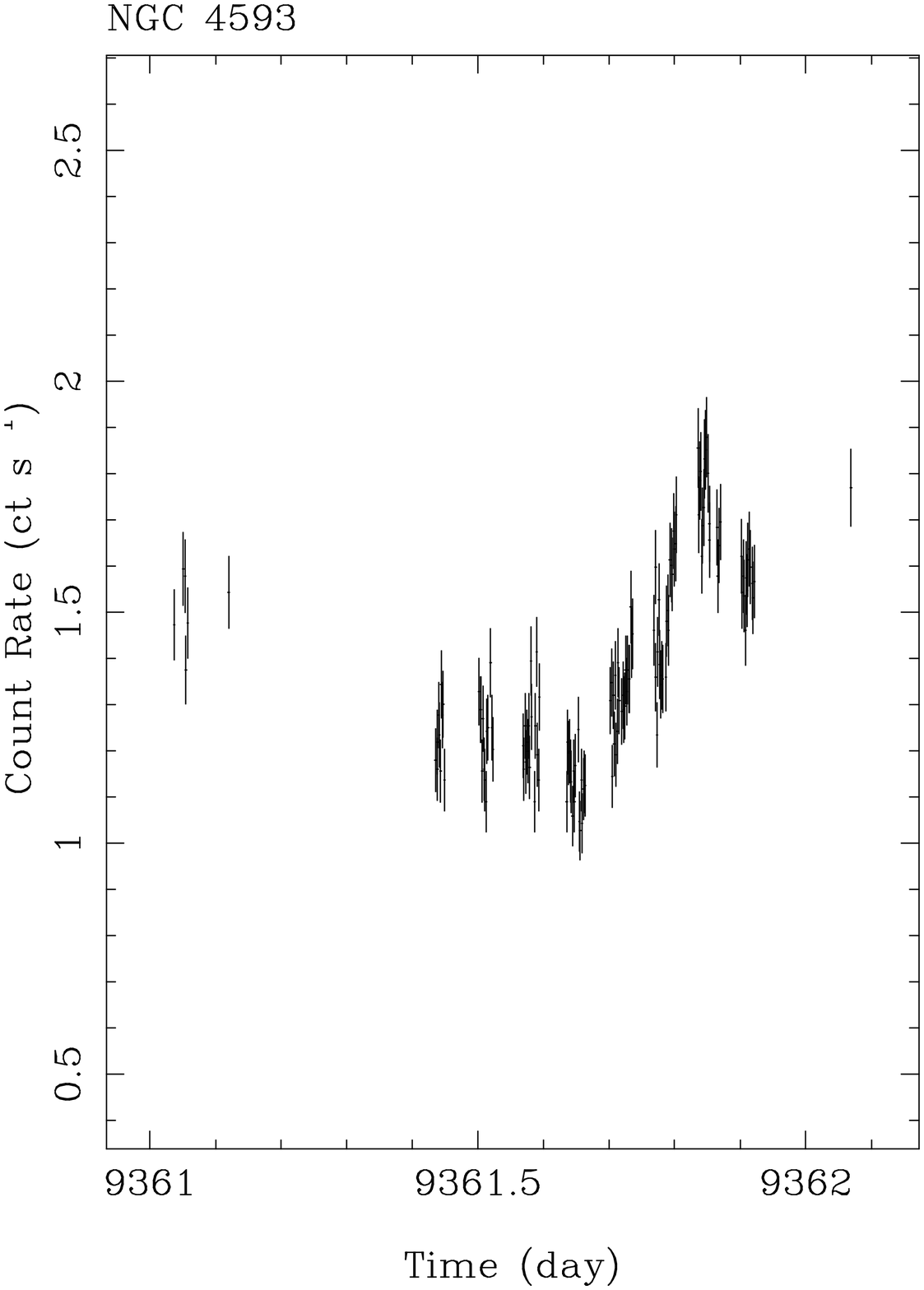}
\epsfysize=0.4\textwidth
\plotone{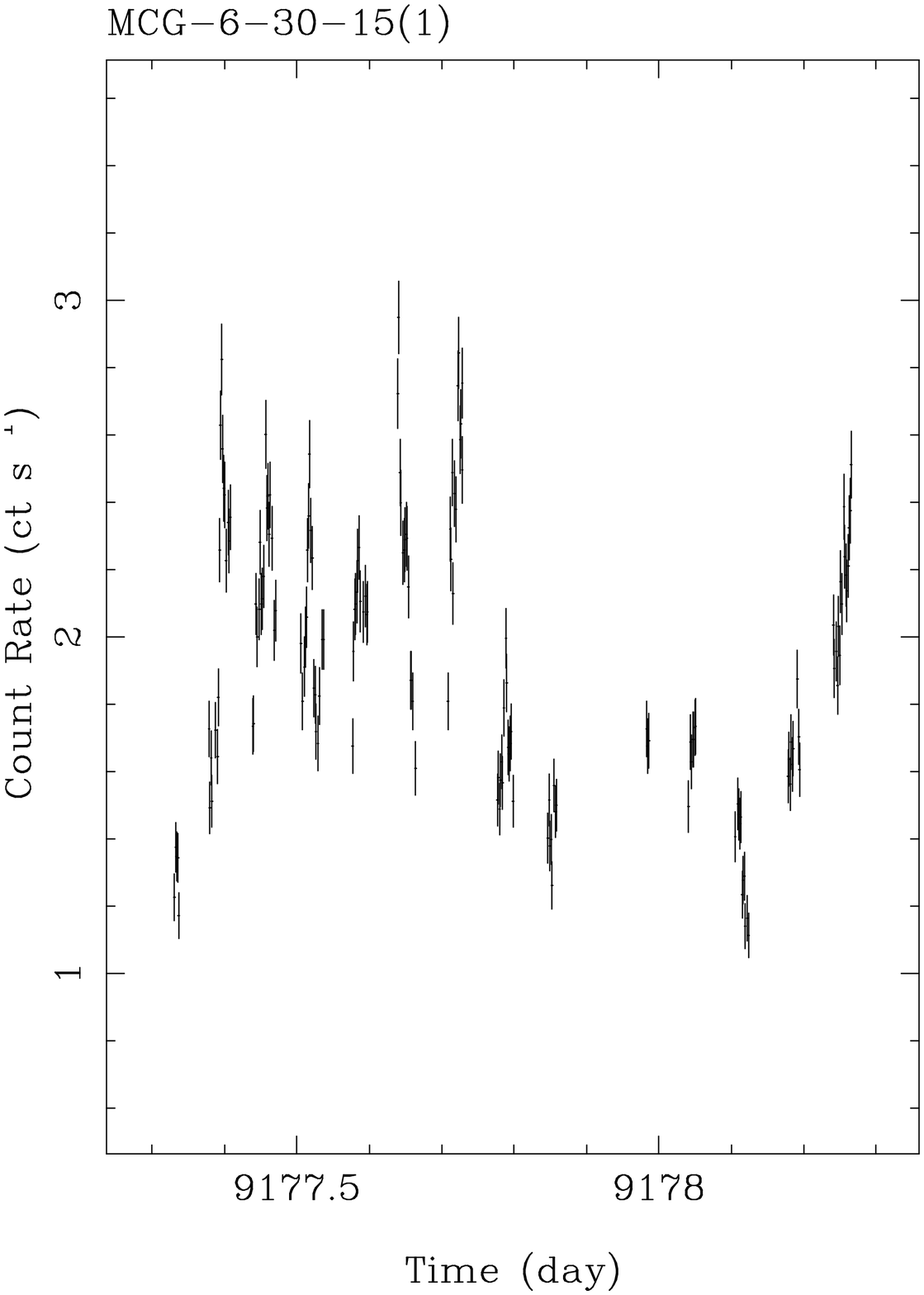}
\end{figure}

\begin{figure}
\epsscale{0.5}
\epsfysize=0.4\textwidth
\plotone{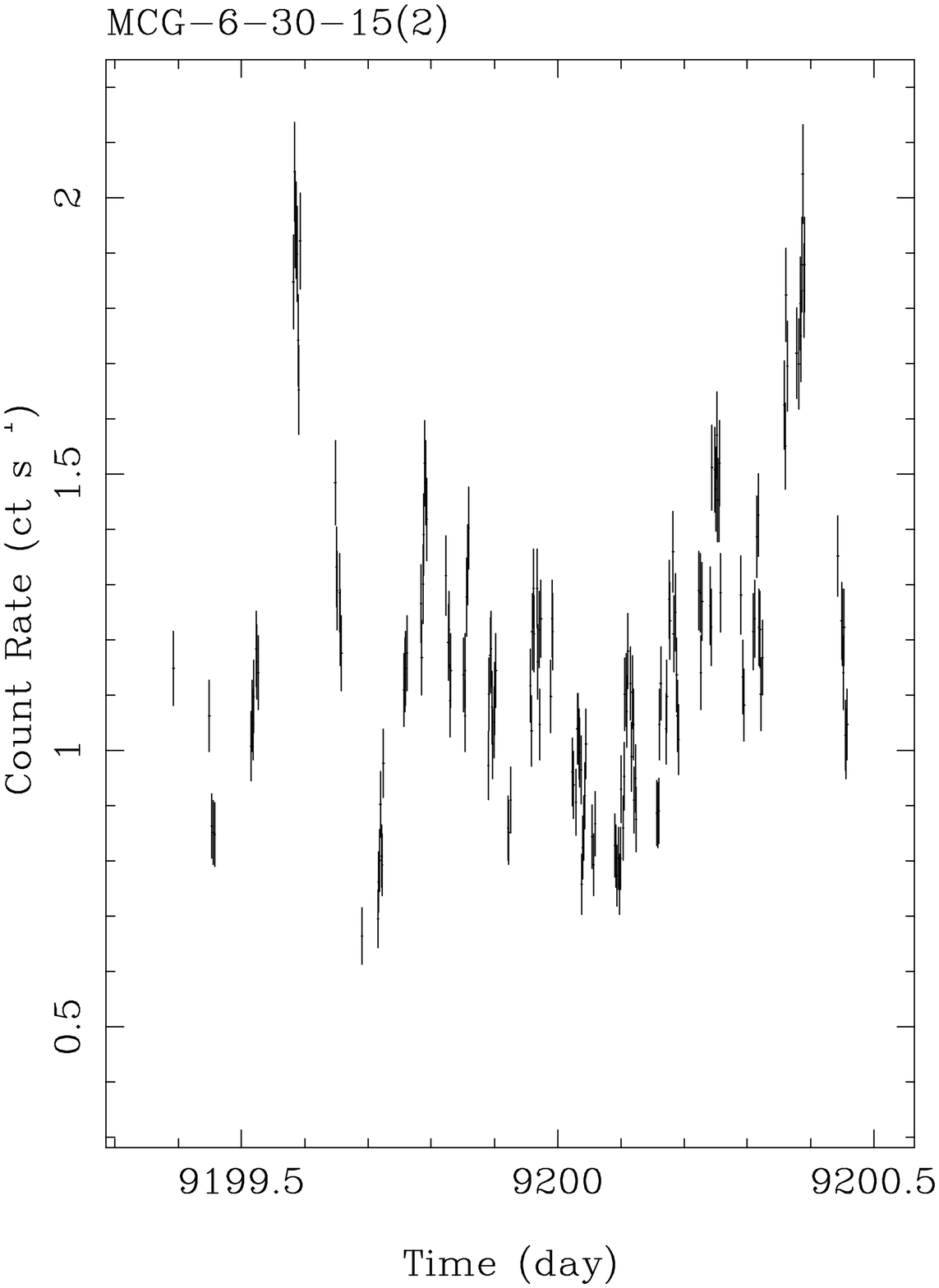}
\epsfysize=0.4\textwidth
\plotone{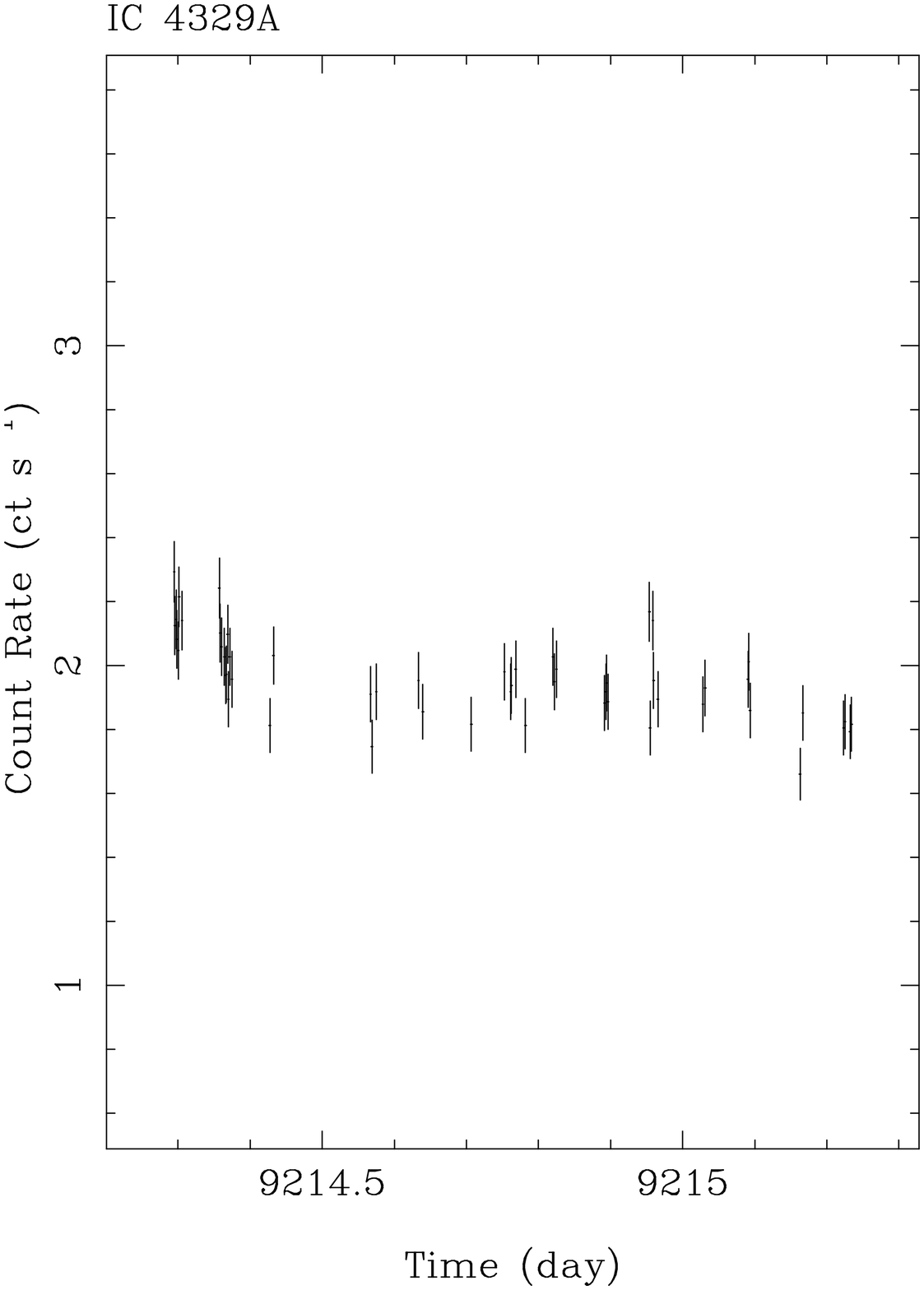}
\end{figure}

\begin{figure}
\epsscale{0.5}
\epsfysize=0.4\textwidth
\plotone{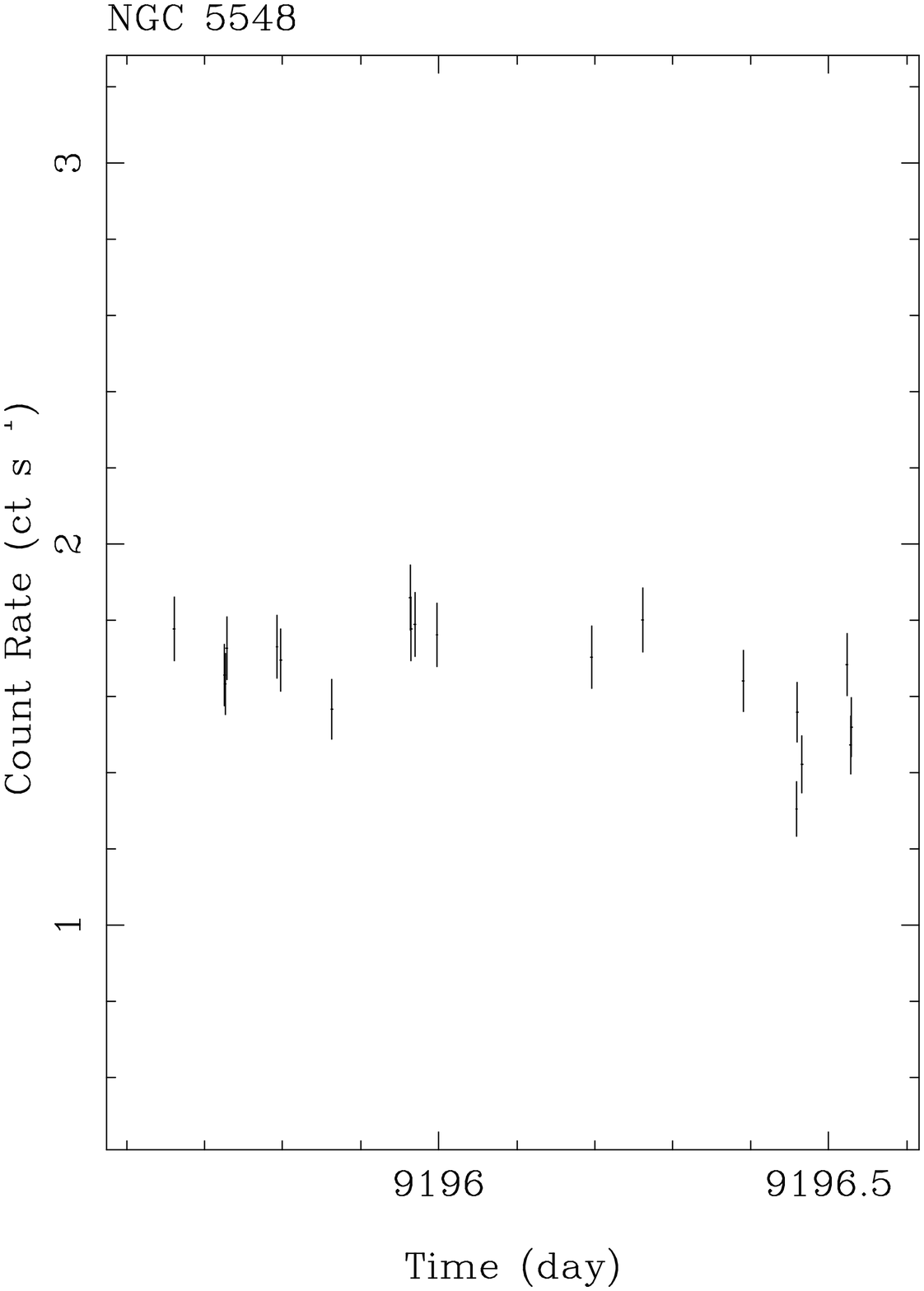}
\epsfysize=0.4\textwidth
\plotone{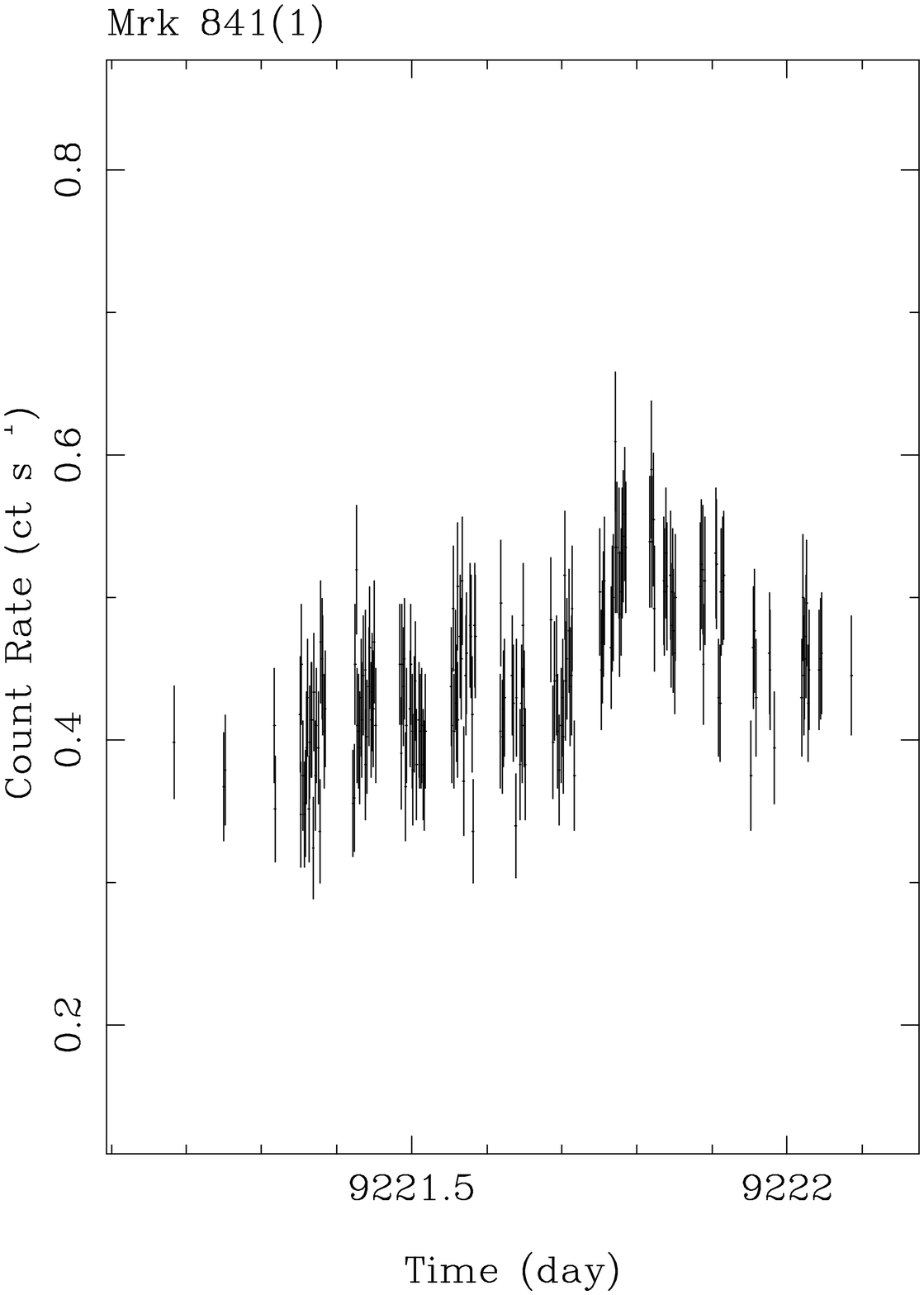}
\end{figure}

\begin{figure}
\epsscale{0.5}
\epsfysize=0.4\textwidth
\plotone{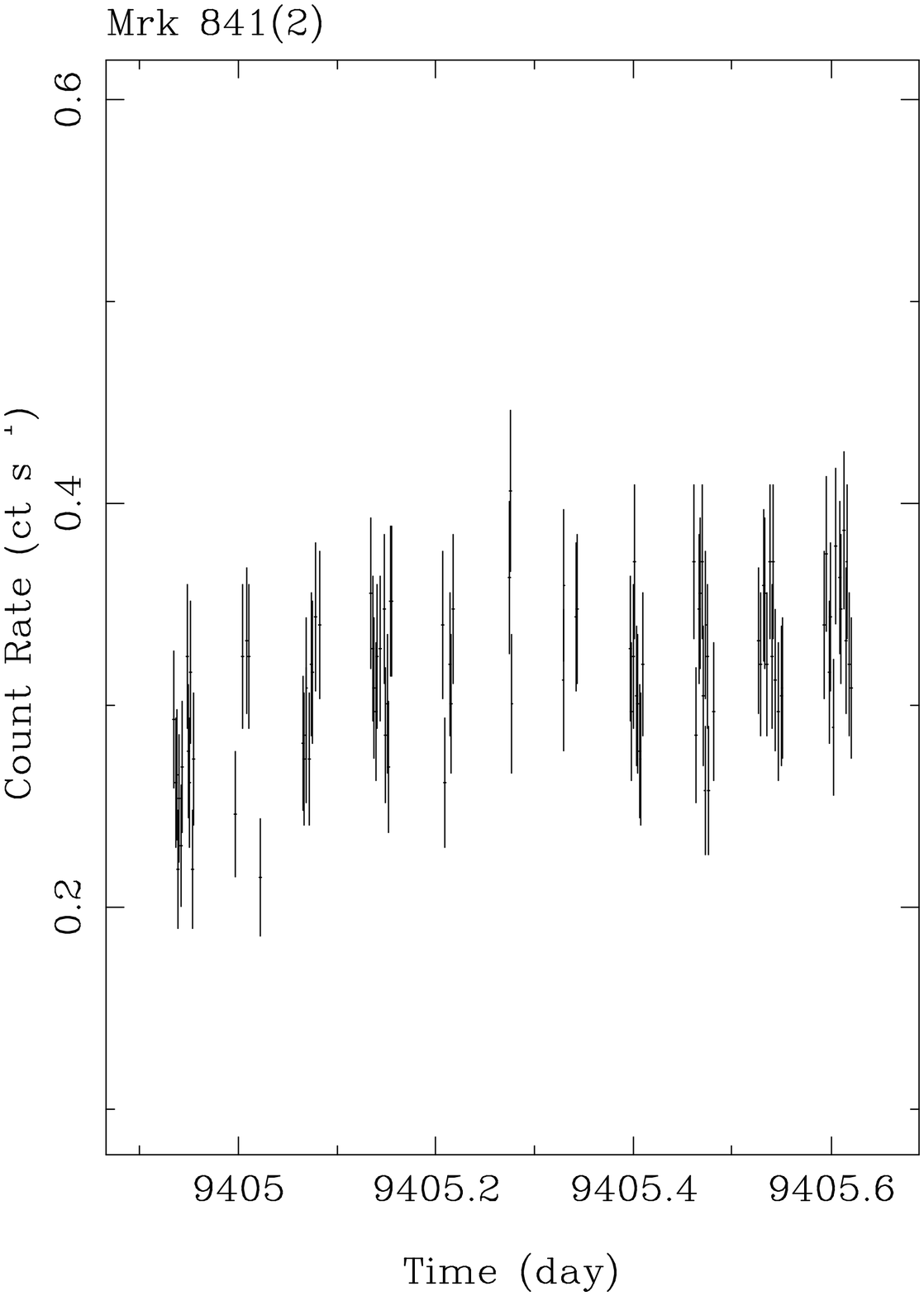}
\epsfysize=0.4\textwidth
\plotone{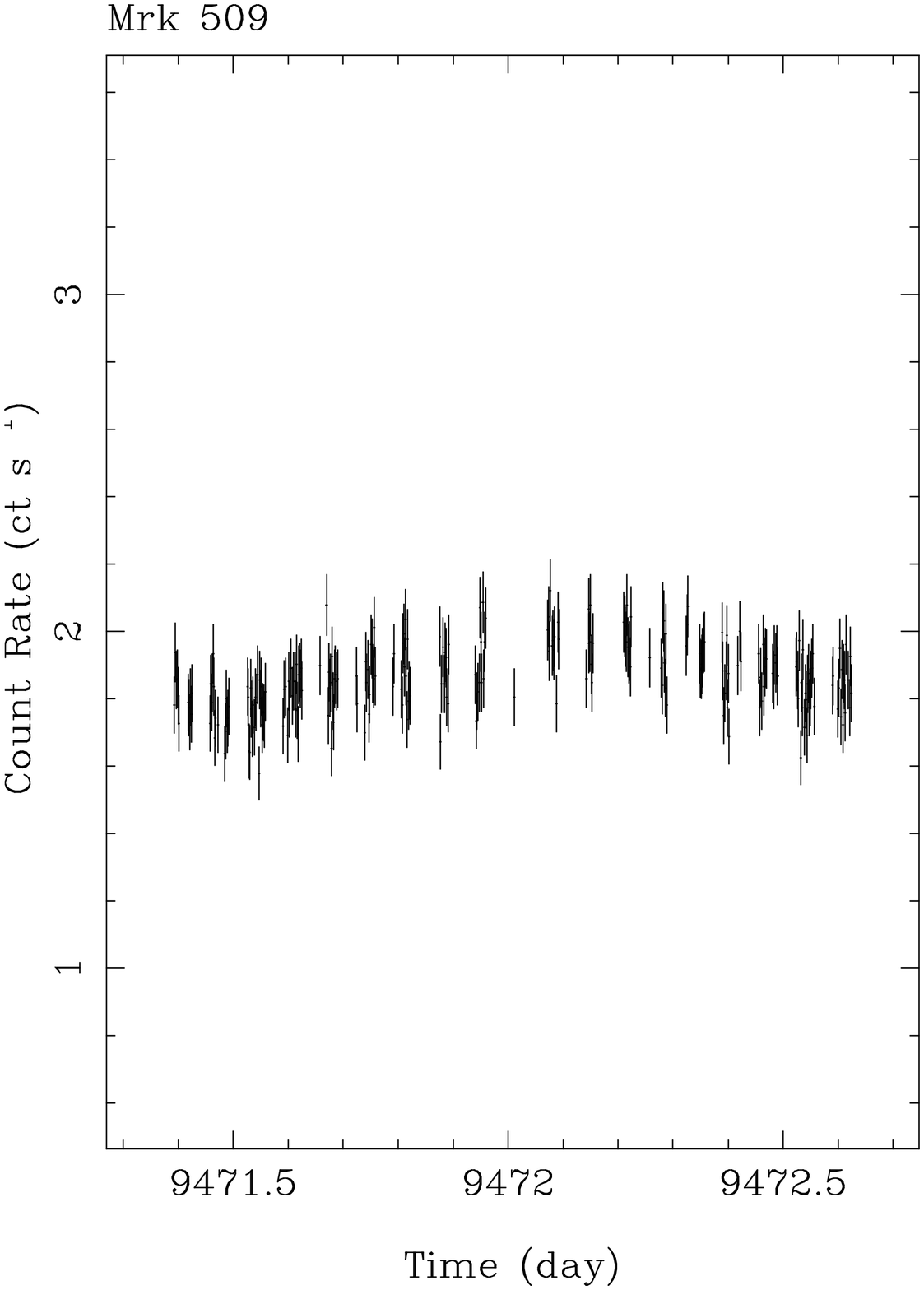}
\end{figure}

\end{document}